\documentclass[pdflatex,sn-basic]{sn-jnl}%
\usepackage{graphicx}%
\usepackage{multirow}%
\usepackage{amsmath,amssymb,amsfonts}%
\usepackage{amsthm}%
\usepackage{mathrsfs}%
\usepackage[mathscr]{eucal}%
\usepackage[title]{appendix}%
\usepackage{xcolor}%
\usepackage{textcomp}%
\usepackage{manyfoot}%
\usepackage{booktabs}%
\usepackage[utf8]{inputenc}
\usepackage{csquotes}
\usepackage{url}%
\usepackage{algorithm}%
\usepackage{algorithmicx}%
\usepackage{algpseudocode}%
\usepackage{listings}%
\usepackage{natbib}
\usepackage{comment}
\usepackage[utf8]{inputenc}
\usepackage{amssymb}
\usepackage{adjustbox}
\usepackage{fix-cm}
\usepackage{lscape}
\usepackage{colortbl}
\usepackage{array}                     
\usepackage{makecell} 
\usepackage{hyperref}
\usepackage{chngcntr}
\theoremstyle{plain}

\DeclareUnicodeCharacter{2063}{}

\begin{document}

\title[Article Title]{%
  Can Large Language Models Understand As Well As Apply Patent Regulations to Pass a Hands-On  Patent Attorney Test?
}

\author{Bhakti Khera, Rezvan Alamian, Pascal A.\ Scherz, and\\ Stephan M.\ Goetz}

\abstract{
The legal field already uses various large language models (LLMs) in actual applications, but their quantitative performance and reasons for it are underexplored. We evaluated several open-source and proprietary LLMs—including GPT-series, Anthropic, Deepseek and Llama-3, 
variants—on parts of the European Qualifying Examination (EQE) for future European Patent Attorneys.

OpenAI o1 led with $0.82$ accuracy and $0.81$ F1 score, whereas (Amazon Web Services) AWS Llama 3.1 8B lagged behind at $0.50$ accuracy, and a Python-deployed Llama 3.1 8B scored $0.55$. The latter two are within the range of mere guessing for the two-answer forced-choice design. None of the evaluated models could have passed the examination fully, as accuracy never exceeded the average threshold of $0.90$ required for professional-level standards---also not models that are regularly promoted for their assumed beyond-PhD- and bar-admitted-lawyer-level performance. GPT-4o excelled at integrating text and graphics, while Claude 3 Opus often lost formatting coherence.

Human patent experts evaluated the textual justifications and uncovered various characteristic traits as well as critical shortcomings of each model. They valued clarity and legal rationale over the raw correctness of the answers, which revealed misalignment between automatic metrics and expert judgment. Model outputs were sensitive to modest temperature changes and prompt wording, which underscores the remaining necessity of expert oversight. Future work should target logical consistency, robust multimodality, and adaptive prompting to approach human-level patent proficiency. In summary, despite the outstanding performance of recent large models, the general public might overestimate their performance. The field has a long way to go to develop a virtual patent attorney. This paper wants to point out several specific limitations that need solutions.

}


\maketitle

\section{Introduction}\label{sec1}

Recent advances in generative artificial intelligence have massively changed natural language processing (NLP). Large language models (LLMs), such as generative pre-trained transformers (GPT), have grown in size and started demonstrating unexpected functionalities as well as performance levels, which rapidly catapulted LLMs into the centre of attention of all deep learning. Despite the behaviour of these models in general NLP and demo tasks, there are notably fewer studies on more practical work in specific domains, such as patent data \citep{jiang2025natural,jiang2024can}.
On the one hand, patents are a legal instrument for protecting intellectual property. On the other hand, the patent literature is mostly text-based and constitutes one of the largest and most formal collections of technological knowledge. Its language is highly technical, with a legally and technically precise language and high requirements for logical stringency. The patent literature, accordingly, may appear as a near-perfect application for LLMs. However, the patent language is very different from everyday language to which LLMs are trained. The language complexity of patents makes them particularly challenging for automation and inaccessible even to technology experts. The European Patent Office (EPO) conducts a rather practical examination\footnote{European qualifying examination (EQE):\\ \url{https://www.epo.org/en/learning/eqe-epac/european-qualifying-examination-eqe}} of patent attorneys, most of whom have already practiced as patent attorneys in their country before, to evaluate their hands-on fitness in the regulations and their typical attorney work to represent clients at the EPO. The exam accordingly represents the practical skills and legal knowledge that patent attorneys should command. If LLMs are to extract and exploit the knowledge stored in the patent literature or help as agents as well as in fully automated systems, models should also be able to answer these benchmark questions and perform these tasks.

Our hypothesis is that open-source LLMs can solve the problems of EPO's qualifying exam under certain conditions and achieve performance comparable to larger, proprietary models. Specifically, we hypothesize the following:

\begin{enumerate}
 \item \textbf{Temperature Variability:}\\
 Adjusting the temperature parameter in larger models such as Llama 3.1 405B, and Claude Sonnet 3.5 and smaller models such as Mistral 7B, and Llama 3.1 8B may influence the language and contextual richness of the responses with or without altering the correctness of the final answers (\textit{True} or \textit{False}). Thus, temperature tuning could optimize response quality without compromising accuracy.
 \item \textbf{Prompting Techniques:}\\
 The use of minimal prompting versus no prompting in models such as Llama 3.1 405B, Claude Sonnet 3.5, Llama 3.1 8B, and Mistral 7B may affect the context length and linguistic style, but may or may not affect the correctness of the final answers. Even smaller models may perform effectively with appropriate prompting strategies.
\item \textbf{Model Comparison:}\\
 Among various models tested—including Anthropic, Mistral AI, Google Gemma-9B, Llama models, Deepseek models, and GPT variants, we anticipate that Llama 3.1 405B will provide more precise and detailed answers. Initial anecdotal observations suggest that OpenAI o1 outperforms other models. It would indicate a potential for optimized open-source models in specialized tasks.
\item \textbf{Context Length Management:}\\
 The length of the input context can impact model performance. In Python scripting, processing each question individually allows for longer context lengths, which could lead to more comprehensive answers compared to processing all questions simultaneously in AWS. 
\item \textbf{Platform Differences: }\\
There may be variations in model outputs when deployed on different platforms (e.g., Amazon Web Services (AWS) versus local Python scripts using Hugging Face models), even when using the same model architecture, such as Llama 3.1 8B. Preliminary results indicate that AWS-deployed models may provide different, even more accurate answers, which warrants further investigation. We speculated that AWS might be running a highly optimized Llama 3.1 8B whereas Hugging Face could be running on a less optimized, raw Llama 3.1 8B. There might be differences in the infrastructure and inference speed. It could lead to differences in the output.
\item \textbf{Multimodal Capabilities: }\\
 Testing the models' ability to handle multimodal inputs, such as PDF-to-text conversions for complete exams (Parts 1--2), may reveal additional strengths or limitations in automating the examination process. We tested the multimodal models GPT-4o and Opus.
\item \textbf{Question-Specific Performance: }\\
 An analysis that questions yield the most correct or incorrect answers across models can provide insights into the models' strengths and weaknesses and inform future improvements.
\end{enumerate}

We aim to determine whether open-source LLMs, under optimized conditions, which involve temperature settings, prompting techniques, and context management, can match or even surpass the performance of larger and typically proprietary models in the patent domain. This analysis also assesses whether LLMs are capable of performing at a level comparable to human candidates in EPO's exams.

We analyse open-source models, including GPT-based architectures, in their performance with the rather practical tasks and questions in the exam. We evaluated these models on the basis of the correctness and explanation of their answers. We use established metrics, such as F1 score, accuracy, and language evaluation scores, including Bilingual
Evaluation Understudy (BLEU), Recall-Oriented Understudy for Gisting Evaluation (ROUGE), and BERTScore. Human evaluation by patent professionals assessed the quality of each model's explanations and justifications. Word embedding similarities in open-source models provide crucial information on how models struggle with legal word distinctions.

Our findings demonstrate that these open-source models can achieve respectable results under certain conditions and have the potential to automate aspects of the patent process. By comparing the performance of various models and testing different parameters, we provide a comprehensive analysis of the capabilities and limitations of LLMs in this specialized domain.

This paper wants to shift the focus from general-purpose NLP applications to the specific demands of the patent literature and thus contributes to the development of more efficient and robust tools in intellectual property management. Our work not only answers the question of the feasibility of LLMs for patent process automation but also identifies critical factors that influence their performance.

\section{Background}\label{sec2}

LLMs represent a major milestone in artificial intelligence and NLP. These models exploit the well-known concept of attention and are pre-trained on vast amounts of text data. Beyond a certain level of model size, they analyze and generate text so well that they excel in both natural language understanding (NLU) and natural language generation (NLG) tasks. The evolution of LLMs reflects broader advancements in NLP, which began with statistical models such as n-grams and hidden Markov models in the 1990s and before \citep{YANNAKOUDAKIS1990509, 4766902, miller1994statistical}. The subsequent introduction of neural networks and word embeddings, such as Word2Vec \citep{mikolov2013efficient} and GloVe \citep{pennington2014glove}, boosted the field by capturing semantic relationships between words. The combination of embedding and neural networks, as well as their derivations, set the stage for the development of modern deep learning architectures.

A significant breakthrough came in 2017 with the introduction of transformer-based models by \citet{vaswani2017attention}, in which attention better captures the longer-ranging relationships in syntagmatic and paradigmatic ties between words or tokens and dramatically improves the efficiency as well as accuracy of text processing. Transformers and derivations dominate today’s most prominent LLMs. These include OpenAI’s GPT series \citep{openai2023gpt4}, Google’s BERT \citep{devlin2019bert} and T5 \citep{ni2021sentencet5scalablesentenceencoders}, Meta’s RoBERTa \citep{liu2019robertarobustlyoptimizedbert} and Llama \citep{touvron2023llamaopenefficientfoundation}, as well as Mistral AI's \citep{jiang2023mistral7b} and Anthropic's various models \citep{anthropic2024claude}.

LLMs have demonstrated exceptional versatility across a range of applications. The generative pre-trained transformer (GPT) series, which also encompasses GPT-2 \citep{radford2019unsupervised}, GPT-3 \citep{brown2020language}, and GPT-4 \citep{openai2023gpt4}, has set the standard for high-quality text generation, summarization, translation, and question answering. GPT-3, with its 175 billion parameters, and GPT-4, which scales even further, exemplifies the potential of LLMs in multi-functional NLP tasks. Recent iterations such as GPT-4o and the o1-series aim at further increasing performance.

Encoder-only BERT introduced bidirectional context and focuses mostly on the analysis as well as understanding of language, while encoder--decoder models T5 and BART combine text understanding and generation to unify NLP tasks into a versatile text-to-text framework \citep{raffel2020exploring, lewis2020bart}. PaLM \citep{chowdhery-2022} and its successor Gemini use the decoder-only structure of many recent LLMs with a focus on text generation and offer scalability for diverse tasks, while LaMDA aims at conversational AI to generate meaningful and natural dialogue \citep{thoppilan2022lamda}.
Various models such as Codex particularly specialize in software development and assist with tasks such as code generation and debugging \citep{chen2021evaluating}.

With the growing size of models, corporations started to close the source of models. GPT-2 was available in code and well analyzed by the community; all of OpenAI's later developments remained closed. Other players followed suit. For some time, it appeared the top class of models would be proprietary, inaccessible, and undocumented. Such a development may risk mono- or oligopolizing the technology, close out large parts of the community as the cost for pre-training has reached excessive levels, and also hamper progress.

The Llama series, however, went from restricted access, potentially motivated also by a code leak, to some form of open source. More recently, the open-source DeepSeek series with effective model sizes from 1B to around even 600B parameters was trained from scratch with on the order of a trillion tokens \citep{guo2024deepseek}. The coder version sourced training data from 87 programming languages organized at the repository level, which enables an enhanced cross-file understanding. A major technological breakthrough of the recent versions is the division of the entire model into smaller experts, each with substantially fewer degrees of freedom, which limits the concurrently active model size and appears to drastically reduce both the cost of training and inference despite similar performance to monolithic models \citep{liu2024deepseek, guo2024deepseek}.

Despite their capabilities, LLMs face several limitations. They rely heavily on large-scale internet data, which typically includes inaccuracies, biases, and incomplete information as well as poor language, often colloquial, imprecise, empty, and grammatically questionable. As a result, they can propagate misinformation and exhibit biased behavior, particularly in applications requiring domain-specific accuracy, such as medicine or law \citep{bender2021stochastic,weidinger2021ethical}. While LLMs excel in generating human-like text, they often struggle with complex queries requiring multi-step reasoning or deep contextual understanding \citep{cobbe2021training}. These challenges are compounded by the significant computational resources required to train and deploy such models, especially those with billions of parameters.

The application of deep learning and LLMs to the patent domain is an emerging field. Recent surveys provided a summary of the applications \citep{jiang2025natural}. Despite intensive research on patent summarization \citep{sharma2019bigpatent} and translation \citep{pouliquen2015patent}, there is a noticeable gap in research on the generative and analytical use of LLMs.
Few studies tested large language and multimodal models for claim generation or for writing descriptions in part or in full \cite{knappich2024pap2pat, wang2024autopatent, shukla2025patentlmm, jiang2025enriching, chen2024dependency, jiang2024patent}. However, metrics for good quality are limited \cite{trapp2024llm,yoo2025patentscore,chen2025evaluating,jiang2025towards}.
Other work used text-mining, graphical, and general deep-learning methods \citep{abbas2014literature,krestel2021deeplearning, shalaby2019patentretrieval, just2024datascience}.

\citep{lee2020patent} investigated the use of GPT-2 for claim generation and found that minimal training was sufficient to produce patent-like text, i.e., which sounded like a patent. However, the quality of the text appears to be poor. Various others tested GPT-2 and formed domain-specific models
\citep{leettransformers}. Overall, GPT-2 and similarly sized models could be considered substantially outdated or not on par with recent models. 

The ultimate goal appears an automated patent examiner or agent \cite{wang2024autopatent,lee2024instructpatentgpt,wang2024evopat,bui2025advancing, wang2024texttt}. However, it is neither clear if available LLMs can manage the specific language and content requirements of the patent world nor which model might be most appropriate. For sufficient input, these recent LLMs are pretrained on large general corpora, and particularly also a massive body of low-quality online text. Previously, GPT-3 was found to struggle with the actual requirements of patent language \citep{tu2022limits}.

This work aims to fill this gap. We evaluate a number of recent LLMs, closed and particularly also open-source models, and expose them to a domain-specific test to evaluate the performance and understand limitations to inform the development of more appropriate tools.

\section{Methodology}\label{sec4}
\subsection{Data}
As a relatively practical hands-on test, the European Qualifying Examination (EQE) assesses whether a candidate is qualified to practice as a professional representative before the European Patent Office (EPO). It tests proficiency in European patent law, the International Patent Cooperation Treaty (PCT), the Paris Convention, EPO Boards of Appeal case law, and relevant national laws.
We chose the pre-examination of 2012--2024 (except for 2020, which was canceled due to COVID-19), which tests foundational knowledge on legal aspects (Questions 1--10) and the ability to analyze and interpret patent claims based on provided descriptions and prior art.
\footnote{Pre-examination data: \url{https://www.epo.org/en/learning/eqe-epac/european-qualifying-examination-eqe/compendium/pre-examination}}\footnote{\url{https://github.com/bhaktikhera1/LLM_EQE}}

Each problem contains four questions, and candidates must determine with a written explanation whether a statement is \textit{True} or \textit{False}; in some cases, the answers may require nuanced consideration. We tested the legal part on all models and the claims section, which needs multimodal processing, on a subset of models.
We tested multimodality for both parts of the years 2012--2022.

\subsection{Experiment Design and Models}
The first part of the study evaluates the performance of various LLMs on the legal component of the EQE. Our experiments were designed to assess the models' abilities to understand and argue complex legal scenarios, as well as to explore the impact of different parameters and prompts on their performance.

We tested a range of state-of-the-art LLMs, including both open-source and proprietary models, as follows:
\begin{enumerate}
\item OpenAI GPT-4o  (version made available on August 6, 2024)
\item OpenAI o1 (version o1-preview made available on September 12, 2024)
\item Meta Llama 3.1 405B (AWS version made available on July 25, 2024)
\item Meta Llama 3.1 70B (AWS version made available on July 25, 2024)
\item Meta Llama 3.1 8B (AWS version made available on July 25, 2024)
\item Stand-alone Python Meta Llama 3.1 8B (meta-Llama/Llama-3.1-8B-Instruct from Hugging face, version made available on July 18, 2024)
\item Google Gemma 2 (google/gemma-2-9b from Hugging face, version made available on June 25, 2024)
\item Mistral Large (AWS version, mistral-large-2402)
\item Anthropic Claude 3 Opus (AWS version v1 made available on February 29, 2024)
\item Anthropic Claude 3 Sonnet (AWS version v1 made available on February 29, 2024)
\item Anthropic Claude 3.5 Sonnet (AWS version v2 made available on October 22, 2024)
\item Deepseek-v3 (version 26.12.2024)
\item Deepseek-v3 R1 (version 20.01.2024)
\item Mistral 7B Instruct (AWS version v0.2 made available on December 13, 2023)
\end{enumerate}

These models vary in size, architecture, and training data and provide a comprehensive overview of current LLM capabilities in the legal domain.

Our study involved five structured experiments that examined how large-language models perform on the exam, with special attention to legal reasoning, multimodal comprehension, and response stability.\\

In Experiment 1, we measured each model’s skill at answering EQE legal questions and justifying its choices. We posed both a single item---prompted with \textit{“Please answer the question with justifications!”}---and a set of ten---using \textit{“Please answer all the questions with justifications!”}. The goal was to see how the models interpreted and applied European patent rules under both light and heavy workloads.\\

In Experiment 2, we tested the multimodal capacity by feeding entire EQE pre-exam PDFs to OpenAI GPT-4o and Anthropic Opus for the papers 2012--2022. This prompting at once revealed how comfortably each model navigated the mix of text, tables, and figures that patent attorneys handle every day.\\

In Experiment 3, we explored temperature effects. We tested four temperature settings (0, 0.3, 0.7, 1) across Mistral 7B, Sonnet 3.5, Llama 3.1 405B, and Llama 3.1 8B. Re-running a random question (one question was choosen randomly and re-used) at each setting showed where the best trade-off lies between accuracy and creative variability.\\

For Experiment 4, we looked at the power of explicit instructions. We compared answers produced with the directive \textit{“Please answer the question with justifications!”}  for the same models as in Experiment 3 intentionally with no additional prompts (only question, no prompt). The results underscored how a single line of prompt engineering can sharpen or, in some cases, muddle legal analysis.\\

In Experiment 5, we checked deployment consistency by running Llama 3.1 8B in two places: a cloud instance on AWS and a standalone Python build on a local server. Feeding identical inputs to both versions verified whether platform choice alters outcomes crucial for anyone who needs reproducible results.\\

As for implementation, Gemma 2 and the local Llama 3.1 8B ran on a Windows server with 512 GB RAM, while GPT models and any task requiring heavy multimodal lifting used Amazon Bedrock or OpenAI’s infrastructure. Unless temperature was the variable under study, every run shared the same default temperature of 1, top-p of 1, top-k of 50, and a 2\,048-token limit. Finally, we had to exclude Questions 11--20, which contain extensive claim sets and diagrams, because some models could not process that much mixed content. The exclusion kept the playing field level across all five experiments. Also, for Experiments 3 and 4, we used the same questions and models to ensure consistency and comparability of results. These models were re-initialized before every run to avoid issues related to trials.  

\subsection{Evaluation Metrics}
To evaluate the performance of the models on the \textit{True} or \textit{False} statements, we used standard classification metrics, accuracy, and F1 score since the ground-truth answers were available. Accuracy measures the proportion of correct predictions, while the F1 score considers both precision and recall as a balanced assessment of the models' performance on binary classification tasks.
For assessing the quality of the justifications or explanations generated by the models, we employed standard text generation evaluation metrics, including BLEU \citep{papineni2002bleu}, ROUGE-1 (R-1), ROUGE-L (R-L) \citep{lin2004rouge}, and BERTScore \citep{zhang2019bertscore}. For context, BLEU, R-1, and R-L are surface-level metrics that measure the overlap of words or sequences between the generated text and reference texts. Specifically, BLEU evaluates the n-gram overlap between the generated text and the reference text. ROUGE-1 and ROUGE-L assess the overlap of unigrams and longest common subsequences, respectively.

We further picked BERTScore as it goes beyond surface-level comparisons by computing the semantic similarity between the generated text and the reference using contextualized embeddings from pre-trained language models. It should allow for a more nuanced evaluation of the generated justifications' semantic content.\\

\textbf{Human Evaluation}\\

While automated metrics provide standardized and reproducible methods for comparing model performance based on lexical overlaps or semantic relationships, they are self-reflexive and do not involve actual legal or factual knowledge. Thus, they may not fully capture the quality and relevance of legal justifications in the patent domain. We therefore supplemented our evaluation with human assessments conducted by patent professionals. These evaluations strictly adhered to established examination criteria.\\
The human evaluation focused on the following criteria:\\
\begin{itemize}
    \item[--] Completeness of essential features: Assessment of whether the justifications included all necessary legal aspects pertinent to the question.
    \item[--] Conceptual clarity: Evaluation of the clarity and unambiguity of the language used in the explanations.
    \item[--] Consistency in terminology: Check for uniformity in the use of legal terms throughout the justifications.
    \item[--] Technical correctness: Accuracy of legal reasoning and the correct application of patent law principles.
    \item[--] Overall quality: Aggregate measure that combines all the above criteria.

\end{itemize}

Human rating for the models' responses was performed by patent professionals, who assigned points from $1$ (worst) to $10$ (best) based on how well each one matched an ideal correct justification expected from a patent attorney. The grade was better, closer to correct, clear, and logically consistent. In an effort not to introduce grading bias, examiners were blind to the models, i.e., not told which specific models they were grading, because each model was assigned an arbitrary number. Assessment data were available for 2012--2024, and selections were taken from the years 2012, 2014, and 2017. These specific years were chosen at random in order to ensure adequate representation across diverse temporal frameworks and difficulties.

\section{Results}\label{sec5}

This section reports the comparative performance of the evaluated large language models (LLMs) on the EQE pre-examination under the experimental conditions described in Section~\ref{sec4}. 
Because the task is formulated as \textit{True} or \textit{False} statements, a system that answers at random would achieve an accuracy of approximately 0.5 (0.0503 $\pm$ 0.023 for the 480 questions (120 problems with four questions each) when the \textit{True} or \textit{False} ratio is known and exploited). 
Consequently, all models have to compete with this level.
Performance values must be interpreted relative to this baseline and to the class imbalance in the ground-truth labels (46.05\% \textit{True}, 53.95\% \textit{False} in the solution).

\begin{figure}[H]
    \centering
    \includegraphics[width=\textwidth]{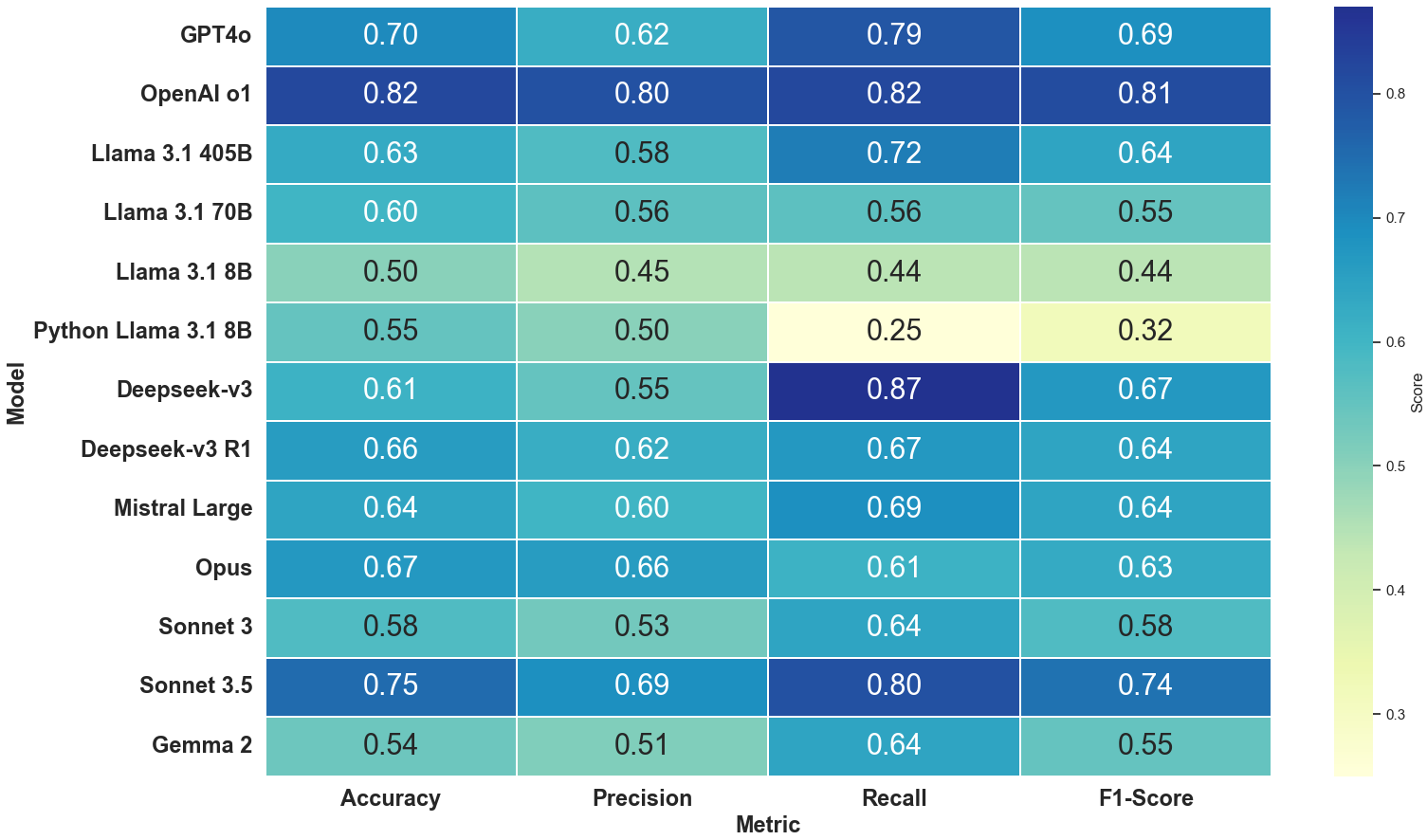} 
    \caption{Heatmap of model performance metrics across different models. Brighter colors signify higher performance values.}
    \label{fig:heatmap}
\end{figure}

\begin{figure}[H]
    \centering
    \includegraphics[width=\textwidth]{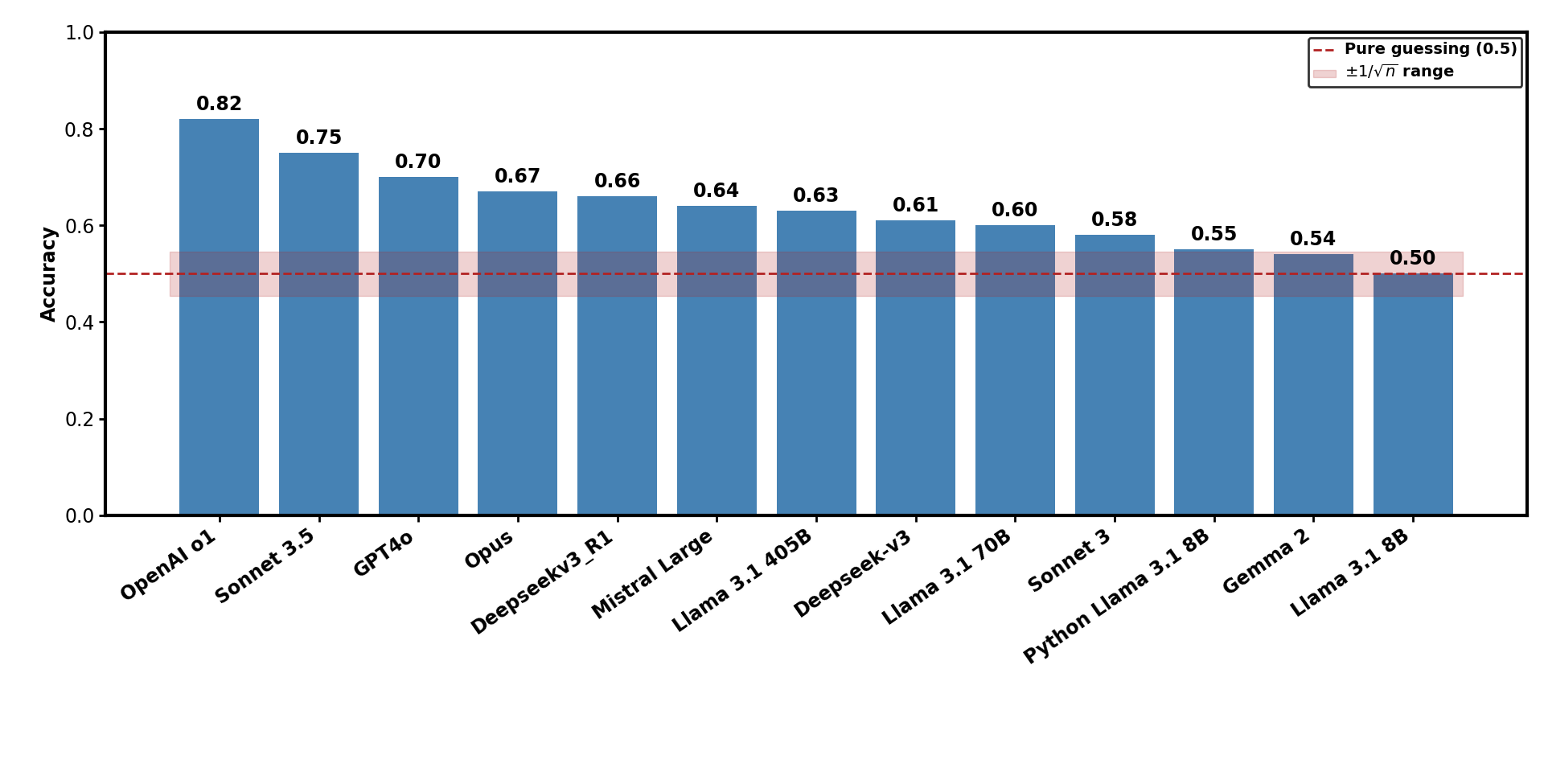} 
    \caption{Accuracy of each LLM on the EQE pre-exam; blue bars show per-model accuracy, the red dashed line marks random guessing (0.5), and the pink band indicates the $\pm 1/\sqrt{n}$ fluctuation range discussed above.}
    \label{fig:heatmaptobarplot}
\end{figure}

Figure~\ref{fig:heatmap} summarises accuracy, precision, recall, and F1 score for each model. {Python Llama 3.1 8B} and more so {AWS Llama 3.1 8B} as well as {Gemma 2} did not surpass the threshold to differentiate their performance from random guessing (Figure~\ref{fig:heatmaptobarplot}).  {OpenAI o1} led with $0.82$ accuracy and $0.84$ F1 score. {DeepSeek-v3} achieved the highest recall ($0.87$) at the expense of precision. Some of that high accuracy appears to reflect a tendency to label a larger
 share of statements as \textit{True}. In contrast, {Python Llama 3.1 8B} attained only 0.25 recall as it in general answered more questions with \textit{False}.

\begin{figure}[H]
    \centering
    \includegraphics[width=\textwidth]{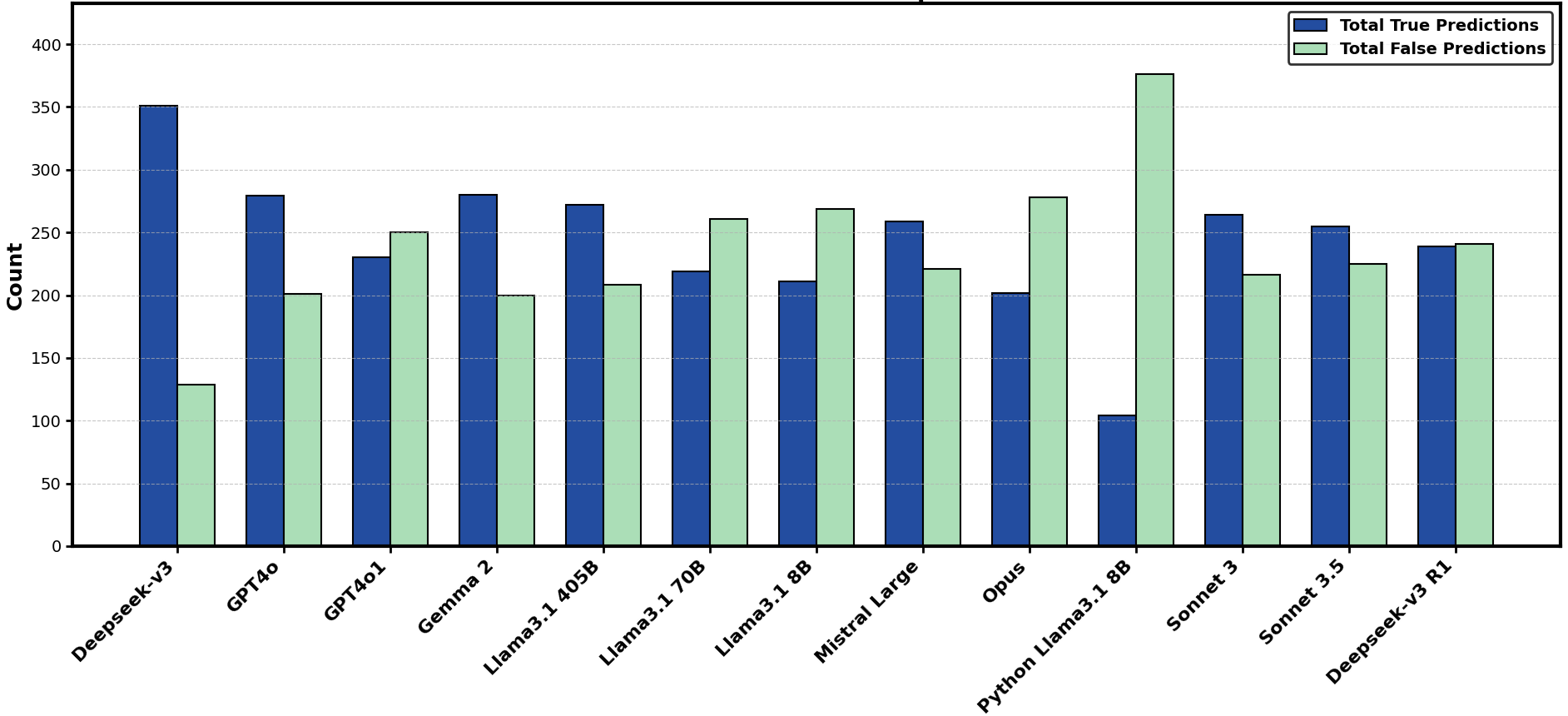} 
    \caption{Distribution of \textit{True} and \textit{False} answers of the various models, notwithstanding their correctness. The various models have a different tendency or bias, which is most pronounced in DeepSeek-v3 and Python Llama3.1 8B.}
    \label{fig:truefalse_bar}
\end{figure}

For those models close to the guessing performance, the small imbalance of the correct answers between \textit{True} and \textit{False} can play a stronger role and amplify model answer biases (Figure~\ref{fig:truefalse_bar}). \texttt{Python Llama 3.1 8B} declared $78.33\%$ of items
 as \textit{False}, which widely exceeded the $53.9\%$ \textit{False} rate in the gold standard; only $21.67\%$ were predicted \textit{True}. This class bias, however, inflated precision for the dominant class,
 but drove recall and therefore the F1 score down. By contrast, the AWS-hosted \texttt{Llama 3.1 8B} produced a near-balanced distribution ($43.95\%$ \textit{True}, $56.05\%$ \textit{False}) and consequently achieved
 noticeably higher recall (0.44) and F1 (0.44) despite substantially lower accuracy. Thus, raw accuracy is insufficient in our setting: one model apparently outperformed the other due to its bias, while both could not exceed mere guessing levels.

Across the remaining models, accuracy and F1 score varied less than recall did. This pattern suggests that individual systems trade precision against recall in different ways while maintaining a broadly similar overall error rate.
 Recall, therefore, provides the greatest diagnostic contrast and should be reported alongside accuracy in future studies.

Patent professionals preferred explanations generated by the AWS-hosted \texttt{Llama 3.1 8B} over those of \texttt{Python~Llama 3.1 8B}, which appeared to conflict with the numerical analysis Fig.\ (\ref{fig:human_scores}). They noted that balanced predictions
 and transparent justifications were more helpful than sparse but highly certain answers. The misalignment between human preference and accuracy underscores the need for composite evaluation and better metrics.

\subsection{Temperature Variability and Prediction Reliability}

\label{sec:temperature}

The sampling temperature ($T$) emerged as the most influential decoding parameter we investigated. When running experiments across four different models under carefully controlled conditions, we observed that decreasing the temperature from a higher setting ($T=1.0$) to a lower one ($T=0.3$) consistently shifted model behavior from highly creative but unpredictable towards more stable and nearly deterministic outputs. At extremely low temperatures ($T \le 0.1$), the outputs became almost perfectly reproducible but occasionally sacrificed accuracy.

\begin{figure}[H] \centering \includegraphics[width=0.75\linewidth]{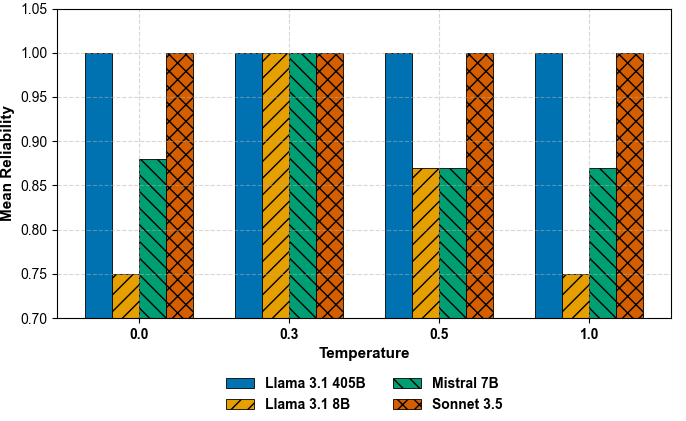} \caption{Mean prediction reliability (mode agreement) as a function of sampling temperature.} \label{fig:reliability} \end{figure}

A particularly optimal performance point, or \textit{sweet spot}, was identified at a temperature of $T=0.3$, especially for the two strongest-performing models, Llama 3.1 405B and Claude Sonnet 3.5. At this temperature setting, both models achieved their highest F1 scores, balancing accuracy and coherence effectively, while also maintaining near-perfect prediction reliability scores (approximately 1.0), as illustrated in Figure~\ref{fig:reliability}.

In contrast, an intermediate temperature setting of $T=0.5$ proved problematic due to increased inconsistency and variability of the output. This instability was particularly evident with models such as Llama 3.1 8B and Mistral 7B. For instance, when we repeatedly tested the same question with four statements at $T=0.5$, we obtained markedly different patterns of answers across multiple trials--specifically patterns such as TFFF, TFTF, FFTF, FTTF, and FTFF (T\,=\,True, F\,=\,False). Such inconsistency led to the prediction reliability dropping noticeably to $0.88$ for Llama 3.1 8B and $0.87$ for Mistral 7B, compared to the perfect reliability score (1.00) observed at $T=0.3$. This increased variance at mid-range temperatures occurs because of the Monte Carlo \citep{metropolis-1949} nature of nucleus sampling \citep{holtzman-2019}, which introduces greater randomness into the response generation as soon as the temperature exceeds zero.

To systematically quantify this observed stability, we applied a resampling technique and introduced the \textit{output reliability metric}. This straightforward metric involves a three-step process: first, repeatedly sampling (bootstrapping) multiple responses for each combination of model, temperature, and question; second, calculating how frequently the most common answer (mode) is reproduced within these samples; and third, averaging this proportion across all questions. A score of exactly $1.0$ indicates completely deterministic model behavior, whereas scores below approximately $0.8$ highlight situations where the model's variability might be too high for safe usage in critical applications. Additionally, through bootstrap sampling \citep{Efron1992}, we calculated 95\% confidence intervals to clearly identify configurations with unacceptable variance during the hyperparameter tuning phase.

From a practical perspective, several clear insights emerged. First, the relationship between temperature and stability is strongly affected by model size: larger models naturally provide more consistent outputs, although they also benefit significantly from moderate temperature reduction (around $T=0.3$). Second, intermediate temperature values (between $0.4$ and $0.6$) consistently present higher risks by significantly increasing variability without delivering meaningful accuracy gains. Lastly, our newly developed reliability metric provides a valuable additional criterion, alongside traditional accuracy and F1 scores, to guide model selection—especially crucial in high-stakes environments such as patent examination.

\subsection{Prompting Techniques}\label{subsec:prompting}

For general results (from all years), we simply used a single line of guidance, “Please answer the question with justifications,” which could improve model performance. 
AWS Llama 3.1 8B gained roughly +2 (from 0.48 to 0.50) percentage points in total accuracy. For these decoders, the extra sentence appears to trigger an explicit chain of thought and eliminated many of the inconsistencies seen in their answers without any prompt beyond the exam questions.
 We also observed that when  AWS Llama 3.1 8B was not given
 any prompt, just the questions, the results were very different. It gave either a one-word response or a response with possible justifications or incomplete responses.

\begin{figure}[H]
  \centering
  \includegraphics[width=.65\linewidth]{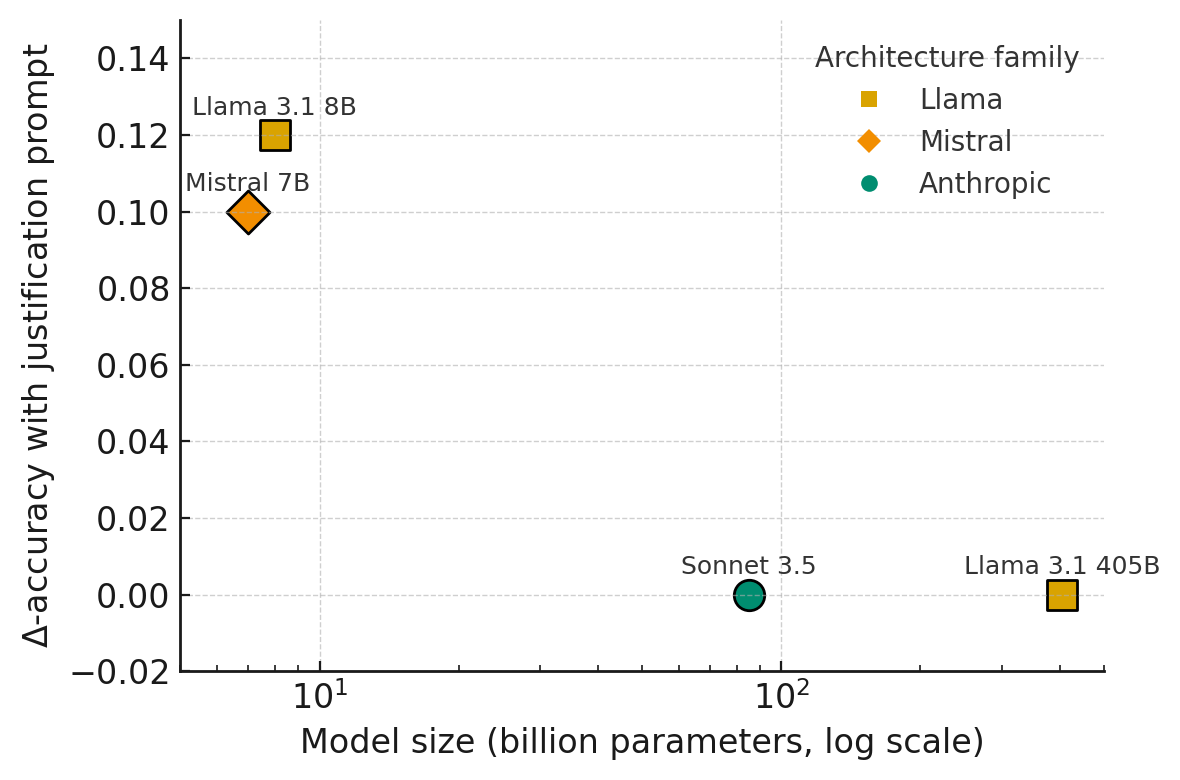}
  \caption{Accuracy improvement obtained by adding an
           explicit justification prompt.}
  \label{fig:prompt_gain_four_correct}
\end{figure}

As a part of Experiment 4, which we only tested on a random question (with four statements) with or without prompts and different temperatures (as stated in Experiment 3), we found that using a prompt vs. no prompt in Mistral 7B not only changed the final answers
 but also the justifications. Llama 3.1 8B and Mistral 7B, gained roughly +12 percentage points (from 0.45 to 0.57) and +10 percentage points (from 0.55 to 0.65) percentage points in absolute accuracy. As Figure \ref{fig:prompt_gain_four_correct} illustrates, the effect was substantial
 for smaller systems and tapered off as model size grew. By contrast, the mid-sized model Sonnet $3.5$ ($\approx 175$B parameters \citep{abacha2025medecbenchmarkmedicalerror}) did not record any notable changes in the final output,
 but in terms of preciseness and conciseness of the justifications. The justifications were more concise and more on point when the prompt was given. Thus, even well-aligned models can extract additional benefit from an explicit request for explanations. The very large Llama 3.1 405B gave detailed answers with rules when given a prompt. The no/minimal gain for large models suggests that instruction-tuned mega-models already embed a latent reasoning process and therefore have less headroom to benefit from the prompt.

Overall, relative gains decline monotonically with model scale: what small models achieve through a carefully worded prompt would otherwise require an order of magnitude increase in parameters. Beyond raw accuracy, human experts also reported sizable jumps in coherence and legal completeness for AWS Llama 3.1 8B, a modest improvement for Sonnet 3.5, and little visible change for Llama 3.1 405B.

These findings underscore that prompt engineering should be treated as a first-class hyperparameter. In many cases, asking for reasoned answers can close most of the performance gap at a fraction of the computational cost.
 Thus, prompt engineering appears to be an attractive option before resorting to larger, more expensive models.

\subsection{Context-Length Management}

We explored two prompt regimes in our context-length experiment. For the Python builds of stand-alone Python Meta Llama 3.1 8B and Google Gemma 2, 
we looped over the exam and sent one question at a time. Each prompt---including statutory references, the four True\,/\,False statements, and the “Answer with justification"
 instruction---remained below the 2\,048-token truncation limit enforced by the Hugging Face generation call in most of the models.\footnote{Many Hugging Face transformer models have a maximum token limit of 2,048 tokens; see, for example, \url{https://discuss.huggingface.co/t/the-current-text-generation-call-will-exceed-the-models-predefined-maximum-length/70393} and \url{https://huggingface.co/docs/transformers/en/model_doc/llama}} We computed token lengths with the Llama 3.1 8B-instruct tokenizer from Tokenizers (v0.15): a single legal prompt (statutory excerpt + four T/F statements + “Answer with justification’’) averages $\approx 420$ tokens ($\simeq 310$ English words), well inside the 2\,048-token hard cut-off.\footnote{According to the Hugging Face model card for \texttt{meta-llama/Llama-3.1-8B-Instruct}, the model supports a context window of up to 128,000 tokens, with a maximum output length of 2,048 tokens. See \url{https://huggingface.co/meta-llama/Llama-3.1-8B-Instruct}}

For every other model---Meta Llama 3.1 405B, Meta Llama 3.1 70B, Meta Llama 3.1 8B, Mistral Large, Anthropic Claude 3 Sonnet, Anthropic Claude 3.5 Sonnet, OpenAI GPT-4o, Deepseek-v3, and Deepseek-v3 R1—we adopted the more demanding setting: the first ten legal questions were concatenated into a single prompt ($\approx\,1\,700$  tokens). This longer prompt length was obtained by summing each question’s token count after running them all through the same tokenizer, mirroring how a practitioner might supply an entire section of the EQE in one API call.

The multimodal trial increased the level of difficulty: the complete PDF of Parts 1–3 (20 questions plus claim drawings, $\approx\,14\,000$  tokens after optical character recognition with figure captions kept) was passed in one shot to OpenAI GPT-4o and Anthropic Claude 3 Opus. Under these conditions, GPT-4o maintained $0.70$ accuracy, whereas Opus dropped to $0.67$. Thus, the various transformer models and their self-attention designs appear to differ more in cache policy and long-context fine-tuning than in raw parameter count \citep{huang-2024}.

Each transformer layer must cache one key and one value vector (each of size 1 × d\textsubscript{model}) per input token so that later tokens can attend back without recomputing earlier activations. When the total number of tokens exceeds the cache capacity, the oldest entries are evicted and the model “forgets” early context. This phenomenon contributes directly to the accuracy drop.

Furthermore, the Python Llama version could score $0.55$ accuracy for the very same questions when its prompt stayed under 2\,000 tokens. By contrast, the AWS-hosted binary—which stores key-value (KV) pairs in bfloat16 and pins them longer in GPU memory—dropped only by 15 percentage points\footnote{Absolute percentage points (for example, 60\% → 45\% = –15 percentage points, not –25\%) to prevent confusion between absolute and relative changes.} at 1\,700 tokens, means that AWS Llama 3.1 8B's accuracy could have been 0.65 instead of 0.5, which equals the performance of guessing.  This accuracy drop indicates that implementation details (numeric precision and cache eviction policy) matter as much as model weights.\footnote{\url{https://docs.aws.amazon.com/bedrock/latest/userguide/what-is-bedrock.html}}\textsuperscript{,}\footnote{\url{https://github.com/huggingface/transformers/blob/main/src/transformers/models/llama/convert_llama_weights_to_hf.py}, describes bfloat16 streaming and
extended KV retention.} These protocol differences single-question loop versus batched prompt and full-paper PDF—are therefore crucial for interpreting the context-length results and are reported alongside every accuracy number in Figure~\ref{fig:heatmap}.

\subsection{Platform Differences}

To isolate deployment effects, we conducted parallel evaluations of the identical Meta Llama 3.1 8B checkpoint using two different deployment environments. Specifically, we compared an AWS-hosted Meta Llama 3.1 8B endpoint with a local deployment using a Python-based Hugging Face implementation. The AWS endpoint processed the first ten EQE questions as a single combined prompt containing approximately 1,700 tokens. In contrast, the local Python deployment answered each of the ten questions individually, with each prompt consisting of roughly 420 tokens.

Notably, these configurations led to differences in performance: the local Python-based Meta Llama 3.1 8B achieved an accuracy of approximately 0.55 when handling shorter, individual prompts. Conversely, the AWS-based deployment, processing a longer combined prompt, yielded an accuracy of approximately 0.50.

The discrepancy in results is primarily attributed to differences in handling context length and key-value (KV) cache management. KV caching is a mechanism used in transformer-based language models to store computed attention keys and values to increase computational efficiency during inference. In scenarios with longer contexts, older KV pairs may be evicted from the cache so that earlier contextual information gets lost.

Amazon Bedrock, the AWS-hosted service, uses optimized strategies such as storing KV pairs in bfloat16 precision, which halves memory usage compared to 32-bit floats. The lower numerical (floating-point) precision allows Bedrock to retain significantly more tokens (upwards of 6,000 tokens) in the cache before eviction becomes necessary. Thus, it maintains relevant contextual information longer at the cost of practically injecting more numerical noise into the transformer. The local Python implementation, on the other hand, executes the standard Hugging Face model without explicit cache optimization, which likely leads to quicker cache eviction and thus reduced accuracy. These implementation differences underline the critical impact of deployment-specific KV caching strategies on model performance.

\subsection{Multimodal Capabilities}
Our multimodal evaluations tested the ability of GPT-4o and Opus to process complete EQE exam PDFs. GPT-4o demonstrated strong performance in extracting and understanding exam content. It achieved higher accuracy and recall across multiple evaluation years. Opus's accuracy ranged from $0.60$ to $0.79$. Recall peaked at $1.00$ in 2017 for Opus. GPT-4o was able to correctly retrieve information, while it maintained high precision in most cases. Additionally, both GPT-4o and Opus
 achieved a consistently high BERTScore (average above $0.80$). The high BERTScore indicates a strong semantic similarity to reference answers.

GPT-4o finished with a mixed performance profile. Whereas it achieved a higher average accuracy (up to $0.83$ in 2013) than Opus in some years,
 it struggled with maintaining stable recall values. Apparently, GPT-4o could not retrieve the correct information and connect important pieces of information. For instance, GPT-4o's recall in 2018 was significantly lower ($0.65$) compared to Opus’s $0.93$. Moreover, Opus' Rouge-L and BLEU scores were lower overall than GPT-4o. Thus, Opus tends to have weaker text similarity in the structured legal explanations.

One key limitation observed in Opus was its handling of complex formatting and multimodal inputs. In contrast to GPT-4o, which successfully parsed and structured PDF content into coherent responses, Opus exhibited formatting errors and inconsistencies in question structuring. Opus may require additional fine-tuning or multimodal integration enhancements to match GPT-4o’s performance in legal document processing.

Overall, these findings emphasize that multimodal capabilities play a crucial role in real-world patent examination that involves non-textual data.  Whereas GPT-4o could mostly extract relevant information and answer structure, Opus still requires improvements in handling complex legal text and maintaining stable recall rates across diverse examination formats.

Tables~\ref{tab:GPT-4o-results} and \ref{tab:opus-results} list the yearly evaluation metrics for GPT-4o and Opus. The results indicate that GPT-4o in many years outperforms Opus in precision, recall, and F1 score. It appears to balance the retrieval of relevant information and accuracy. Notably, GPT-4o maintains a consistently high BERTScore across years due to a stronger semantic similarity to reference answers. On the other hand, Opus demonstrates higher recall in some years. The higher recall values indicate a tendency to retrieve broader sets of relevant information, though sometimes at the expense of precision. The BLEU and Rouge scores also suggest that GPT-4o produces responses that align more closely with reference outputs in structured legal contexts, whereas Opus exhibits greater variability, particularly in complex legal reasoning tasks. These findings reinforce the importance of robust multimodal capabilities for handling intricate legal documents effectively.

\begin{table}[h!]
\centering
\scriptsize
\caption{GPT-4o Results (bold print where GPT-4o exceeds Opus).\\Note: R-1 is ROUGE-1 and R-L is ROUGE-L}
\label{tab:GPT-4o-results}
\begin{tabular}{lcccccccc}
\toprule
\textbf{Year} & \textbf{Accuracy} & \textbf{Precision} & \textbf{Recall} & \textbf{F1 score} & \textbf{BLEU} & \textbf{R-1} & \textbf{R-L} & \textbf{BERT} \\
\midrule
2012 & \textbf{0.80} & \textbf{0.78} & 0.79          & \textbf{0.80} & \textbf{0.16} & \textbf{0.73} & \textbf{0.30} & \textbf{0.78} \\
2013 & \textbf{0.83} & \textbf{0.82} & \textbf{0.84} & \textbf{0.83} & \textbf{0.16} & \textbf{0.68} & \textbf{0.28} & 0.73          \\
2014 & \textbf{0.65} & \textbf{0.58} & 0.61          & 0.60          & 0.08          & 0.65          & \textbf{0.28} & \textbf{0.84} \\
2015 & \textbf{0.68} & \textbf{0.68} & 0.73          & \textbf{0.71} & \textbf{0.11} & \textbf{0.67} & \textbf{0.27} & 0.85          \\
2016 & 0.68          & \textbf{0.64}          & 0.76          & 0.69          & \textbf{0.09} & 0.64          & 0.26          & \textbf{0.85} \\
2017 & 0.71          & \textbf{0.67} & \textbf{0.69}         & 0.68 & \textbf{0.13} & 0.69          & \textbf{0.30} & \textbf{0.84} \\
2018 & \textbf{0.70} & \textbf{0.65} & 0.65          & 0.65          & \textbf{0.07} & \textbf{0.60} & 0.27          & \textbf{0.84} \\
2019 & 0.71          & \textbf{0.64} & 0.75          & 0.69          & 0.07          & 0.60          & 0.28          & \textbf{0.81} \\
2021 & 0.61          & 0.46          & 0.51          & 0.49          & 0.06          & \textbf{0.54} & \textbf{0.30} & \textbf{0.85} \\
2022 & \textbf{0.63} & \textbf{0.60} & 0.62          & \textbf{0.61} & 0.15          & \textbf{0.68} & 0.36          & \textbf{0.80} \\ 
\bottomrule
\end{tabular}
\end{table}

\begin{table}[h!]
\centering
\scriptsize
\caption{Opus Results (bold print where Opus exceeds GPT-4o). \\Note: R-1 is ROUGE-1 and R-L is ROUGE-L}
\label{tab:opus-results}
\begin{tabular}{lcccccccc}
\toprule
\textbf{Year} & \textbf{Accuracy} & \textbf{Precision} & \textbf{Recall} & \textbf{F1 score} & \textbf{BLEU} & \textbf{R-1} & \textbf{R-L} & \textbf{BERT} \\
\midrule
2012 & 0.67          & 0.60          & \textbf{0.89} & 0.72          & 0.11          & 0.60          & 0.27          & 0.77 \\
2013 & 0.77          & 0.74          & 0.82          & 0.78          & 0.12          & 0.28          & 0.26          & 0.73 \\
2014 & 0.61          & 0.57          & \textbf{0.78} & \textbf{0.65} & \textbf{0.09} & 0.65          & 0.27          & 0.83 \\
2015 & 0.60          & 0.60          & \textbf{0.73} & 0.66          & 0.03          & 0.49          & 0.23          & 0.85 \\
2016 & \textbf{0.79} & \textbf{0.80} & \textbf{0.83} & \textbf{0.81} & 0.05          & 0.58          & \textbf{0.30} & 0.83 \\
2017 & 0.71          & 0.59          & \textbf{1.00} & \textbf{0.74} & 0.05          & 0.54          & 0.27          & 0.82 \\
2018 & 0.61          & 0.59          & \textbf{0.93} & \textbf{0.66} & 0.04          & 0.49          & 0.27 & 0.83 \\
2019 & 0.71          & 0.63          & \textbf{0.85} & \textbf{0.72} & \textbf{0.08} & \textbf{0.63} & \textbf{0.30} & 0.83 \\
2021 & \textbf{0.68} & \textbf{0.55} & \textbf{0.82} & \textbf{0.66} & \textbf{0.01} & 0.41          & 0.27          & 0.83 \\
2022 & 0.60          & 0.52          & \textbf{0.70} & 0.59          & \textbf{0.11} & 0.57          & \textbf{0.40} & 0.79 \\  
\bottomrule
\end{tabular}
\end{table}

\subsection{Year- and Question-Specific Performance}

Across the twelve EQE pre-exams (2012--2024, in Appendix \ref{secA}), the panel-wide mean accuracy climbed from $0.59$ in 2012 
to a peak of $0.77$ in 2013, after which it fluctuated between $0.58$ and $0.69$ (median $0.63$). OpenAI o1 set the yearly benchmark accuracy in ten
 of the twelve papers and averaged $0.82\pm0.07$, while Sonnet 3.5 briefly excelled in 2018 ($0.93$ mean accuracy in 2018) and again
 in 2022 ($0.83$ mean accuracy in 2022).  When you look at the accuracy for each individual statement (not just overall or by year), you can see that the model’s success depends more on the specific content or challenge of each question than on the order in time (chronology\,/\,year) in which the questions appeared.
For example, in the 2024 paper, the four nested statements of Question 1 were solved by none of the thirteen models correctly, whereas only three models could answer all the statements from Question 5 correctly. OpenAI o1 was perfect on Questions $6$, $7$, $9$, and $10$. However, for the first cluster (from Questions $1$ to $5$), OpenAI o1 was right only for question $4$. Hence, there were huge differences in the accuracies of both the clusters. For the first cluster, it was $0.75$ and for the second was $0.95$, hence the total mean accuracy in 2024 for OpenAI was $0.85$. This kind of uneven trend, struggling early, then excelling later, was also observed in Sonnet 3, but not in the other models.

Year-by-year comparisons make these trends even clearer. In 2012, OpenAI o1 achieved an accuracy of $0.80$, while Gemma 2 was much lower at $0.38$. The results from 2013 showed that it was the most straightforward year for the models: OpenAI o1 reached $0.95$ accuracy and Python-Llama 8B  $0.58$. In 2015, the difference in performance between models increased again—OpenAI o1 reached $0.88$, but AWS Llama 3.1 8B scored only $0.38$, which is even less than what you would expect from random guessing. This pattern repeated in 2021, when OpenAI o1 scored $0.75$, while AWS Llama 3.1 8B managed only $0.42$. There were only two years when OpenAI o1 was not the top performer: Sonnet 3.5 outperformed it in 2018 (with an accuracy of $0.93$) and in 2022 (with $0.83$). In 2024, however, OpenAI o1 was again
 the most accurate at $0.83$, while AWS Llama 3.1 8B was at the bottom with $0.40$. These ups and downs in accuracy mostly reflect how many complex, multi-part questions were included each year. The exams in 2015
 and 2021 posed several such challenges. \footnote{In isolation for the 80 questions of 2015 and 2021, an accuracy of $0.59$ is statistically within the tolerance of guessing.} Model performance over the years reflects these yearly trends. OpenAI o1 always stayed above $0.70$, reaching its highest accuracy of $0.95$ in 2013, showing a strong
 and steady track record. GPT-4o scored between $0.60$ 
and $0.78$. It performed best in 2016--2019 but dropped in 2021.
 Sonnet 3.5 ranged from $0.60$ to its highest of $0.93$ in 2018, with an overall average of $0.75$. Opus and Mistral Large usually had accuracies in the mid-$0.60$ range, and each reached about $0.80$ or $0.88$ (Opus in 2018) in their best years. Llama 3.1 405B and 70B models stayed in the low-to-mid $0.70$ range. At the lower end, Gemma 2, Llama 3.1 8B, and Python-Llama 8B stayed below $0.75$ in most years and rarely went above $0.50$ on the hardest questions, which is close to random guessing.

That tasks focused on definitions have become fairly easy for today’s models. However, no system has reached an average accuracy above $0.90$ from start to finish. More complex questions that need conditional reasoning, handling dates, or analyzing information from multiple sources still result in lower accuracy by about one-third. The year-by-year data suggest that to improve further, models will need focused training for these challenges, not just small changes like adjusting the temperature setting.

\subsection{Human Evaluation of Models}

\subsubsection{Scores}

\begin{figure}[H]
  \centering
  \includegraphics[width=\linewidth]{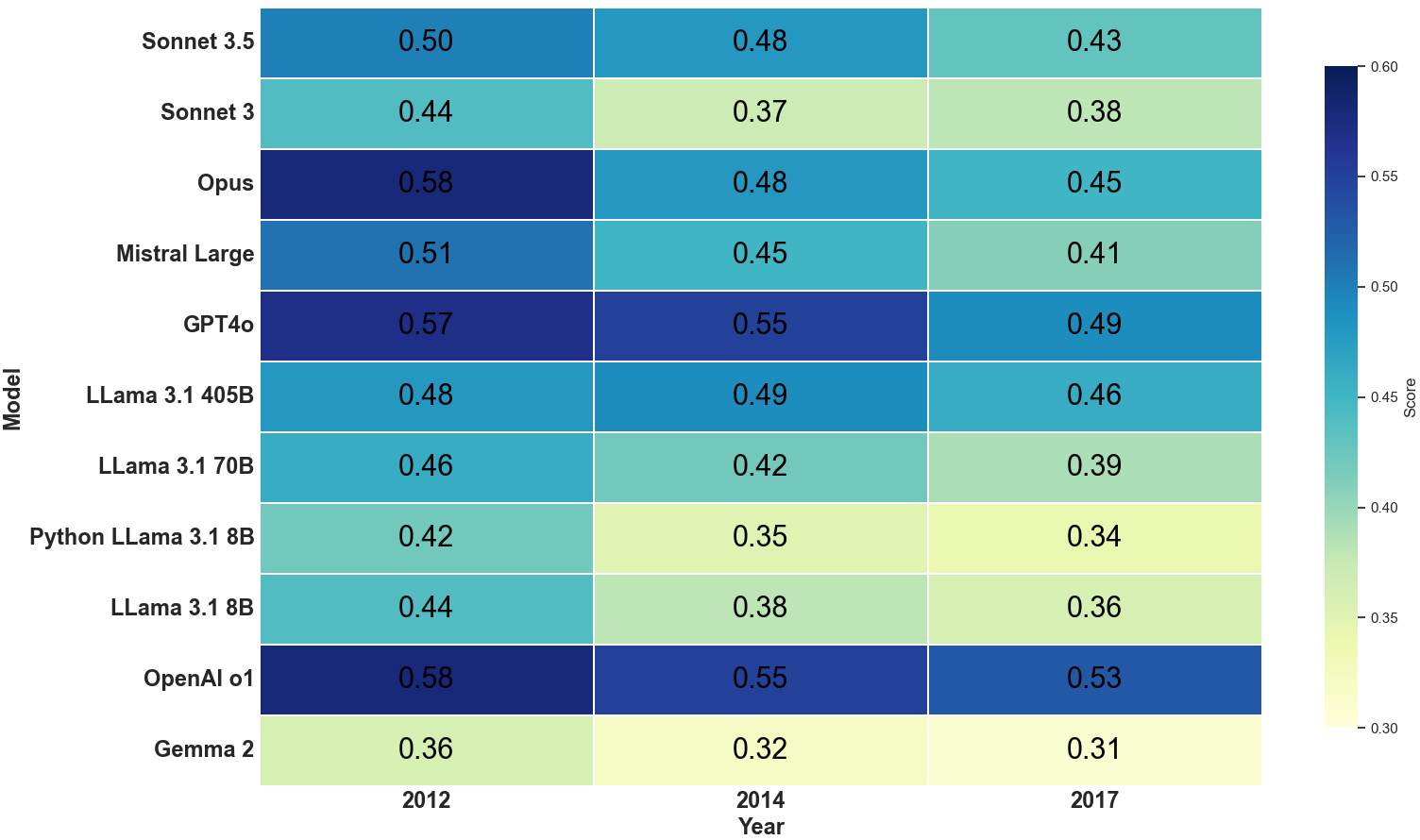}
  \caption{Average human evaluation scores (normalized).}
  \label{fig:human_scores}
\end{figure}

Patent professionals evaluated the models and scored each response on a normalized scale from 0 to 1 (linearly mapped from the original scale of 1 to 10) to facilitate quantitative comparisons. The evaluators were blind to which model generated which response. Figure \ref{fig:human_scores} summarizes the average human evaluation scores across the selected years.

The assessors spent significant time examining these responses and particularly the content in contrast to automated metrics.  They checked reasonableness and logic, as well as the references and citations.

OpenAI o1 consistently received the highest human evaluation scores. In contrast, Gemma 2 consistently performed the worst across all evaluated years. The year 2012 generally presented easier evaluation conditions as models received higher scores on average, whereas the questions from 2017 appeared to be the most challenging. This observation contrasts with the automated evaluation metrics, which rated the results of the earlier exams worse. Thus, the \textit{True} or \textit{False} accuracies may have improved, but the expert assessors found massive flaws in the justifications.

\subsubsection{General Observations Noted by Human Experts}

Across the suite of models, professional evaluators observed a series of systematic qualitative patterns and behaviors that extend significantly beyond raw accuracy alone. 
These observations highlight how models differ in areas such as the computation of deadlines, which are an important aspect in patent cases and can be complicated if deadlines fall on public holidays, rules and statutory citations, legal reasoning, clarity of explanation, and differentiation between novelty and nonobviousness.\footnote{Many if not most of the models could cite the definition, but when specifically asked about a clear case in the questions (e.g., Question 5.4 in 2012), they incorrectly considered a legally novel but obvious (combination of features from different documents from the prior art) invention still not novel. In several cases, the mix-up of novelty and nonobviousness\,/\,inventive step occurred right after an explanation of the difference between both.} Each model displayed its own distinct strengths and limitations, which are described in more detail below.

\paragraph{Sonnet 3.5} This model was consistently noted for producing calculation errors and failing to include relevant rules or articles, e.g., of the European Patent Convention, in its responses. It rarely cited any regulations. 
Evaluators found that its explanations were often incorrect, very brief, and\,/\,or that the model demonstrated a clear lack of \textit{understanding} of case law and guidelines. Still, the model may be among the minority of models that \textit{understood} the difference between novelty and nonobviousness and could apply it correctly, though its explanations in these cases were so short that a decisive decision if Sonnet can reliably employ this most elementary property appears not possible yet. Overall, the issues suggest foundational shortcomings in both practical legal skills and (date) arithmetic as well as reasoning capabilities. 
 
\paragraph{Sonnet 3} On the one hand, Sonnet 3 in contrast to version 3.5 routinely cited legal references in its answers. On the other hand, a considerable share of them was wrong.  Furthermore, it provided explanations that were not only incorrect but also lacking in clarity and substantiation. Thus, Sonnet 3 may attempt to provide a structured legal reasoning; the underlying content is often inaccurate, and the results rather imitate a clear argument.

\paragraph{Opus} The Opus model cited regulations only sporadically and often lacked clear explanations. Yet, it showed potential strength in distinguishing 
between the different legal meanings of novelty and nonobviousness, where the majority of the models struggled, although this observation came with some uncertainty. Interestingly, Opus was further one of the few models that correctly accounted for weekends in the calculation of deadlines (e.g., Questions 2012--Q10.2 and Q10.4). On the other hand, it missed public holidays in other questions, specifically Easter, although the question fairly mentioned Easter to exam candidates. Opus also failed to calculate some other deadlines, e.g., because it used the wrong reference date from the given information. The ability to deal with the high complexity of deadlines in the case of weekends but failing even some simpler calculations may suggest a degree of procedural awareness, albeit inconsistently applied, or some level of chance.

\paragraph{Mistral Large} The patent professionals noted that Mistral Large struggled with dates. It correctly extracted dates but showed tendencies to fail in their interpretation and generated confusion or weird explanations, which were sometimes causally wrong but accompanied a correct overall answer based on the wrong argument. The model 
also rarely included references to the legal regulations. Nonetheless, it was also one of the few models that demonstrated a nuanced understanding 
of the distinction between novelty and nonobviousness. It successfully applied this distinction and correctly identified an invention where only some of the features were in a published prior-art document and the rest in an earlier application of which the case in question claimed priority. The correct handling of those essential properties and the application to concrete cases indicate partial strength in substantive legal reasoning, despite weaknesses in procedural accuracy.

\paragraph{GPT-4o} GPT-4o was criticized for providing poor and inaccurate explanations and for barely including any references at all. In the rare cases, when it cited references, potentially stimulated by questions that also included some, they were often wrong and not even close.
Interestingly, the model surprised the evaluators several times, when it failed in earlier sub-questions but \textit{started to understand} the situation in a later sub-question that built on top. Its performance was perceived as inconsistent and explanations were sometimes called outright poor. Overall, the model may demonstrate some skills in picking up causal chains correctly and drawing proper conclusions, even though the explanations are typically not too useful, which suggests some latent reasoning capabilities that do not reliably surface.

Although GPT-4o could provide the definition of novelty and inventiveness\,/\,nonobviousness in its explanations to questions and even applied it in a few simple examples, it still failed when the case included temporal information in the form of dates or priorities, and was needed for sorting out which document preceded which.

\paragraph{Meta Llama 3.1 405B} According to the blind human experts, this model was one of those that stood out positively among its peers. It was noted for providing references in many cases. Furthermore, it was a model that correctly differentiated between novelty and nonobviousness. It explained this distinction accurately and applied it correctly in several cases. However, Llama 3.1 405B tripped over priorities and dates in other cases or wrongly assigned given features to prior documents so that in itself correct arguments were based on wrong assumptions of features. 
 Evaluators highlighted the model's proficiency with calculations and its ability to evaluate legal deadlines effectively. However, it was not without flaws: in some instances, the model still offered incorrect explanations, explained correctly but derived the wrong conclusion from the explanation, or omitted references.

\paragraph{Meta Llama 3.1 70B} In contrast, the smaller Llama 3.1 70B model rarely provided references and consistently struggled with differentiating obviousness from novelty, which was a common issue across most models. It encountered problems extracting and interpreting dates from the text. It furthermore struggled with calculating deadlines. These weaknesses indicate challenges with both substantive and procedural legal reasoning.

\paragraph{Python Llama 3.1 8B} The Python implementation of Llama 3.1 8B as one of the smallest models in the comparison attempted to include citations—sometimes in large numbers—in various formats (e.g., just reference numbers in square brackets, sometimes with, sometimes without a corresponding bibliography, specific rule numbers, website addresses, etc.). However, most of these references were either fabricated or incorrect, particularly those referring to the European Patent Convention (EPC) and the Patent Cooperation Treaty (PCT). It failed to distinguish between novelty and obviousness and was unable to interpret
 dates correctly. It sometimes calculated the correct deadline in the explanation, but still disagreed with the correct outcome, which evaluators described as “absurd”. Additionally, the model's explanations were often overly verbose, repetitive, and incorrect, with issues in coherence and legal soundness. The model from time to time referred to specific requirements in EPC articles, which indicates that the respective or related texts were in the training set but potentially overwhelmed the model structure and flexibility

 More critical appears to be that the model sometimes hallucinated wrong facts and added them to the case (including dates that were not given).

\paragraph{AWS Llama 3.1 8B} The answers of the AWS implementation of Llama 3.1 8B contrasted with those of the Python implementation. Whereas the Python version sometimes generated excessive reference lists, the AWS version did not provide any references, was extremely thin-lipped, and was often criticized for incorrect explanations. It frequently missed the logic of the question and had difficulties distinguishing between inventive step and novelty. It also did not grasp the impact of claiming priority so that this priority-providing document does not belong to the critical prior art anymore, which other models did. The model often repeated verbatim legal rules without offering any meaningful 
interpretation, such as understanding that a given date might fulfill a specific deadline. Despite these flaws, it was occasionally able to calculate deadlines correctly. Overall, the model was found to maybe have some limited procedural competence.

\paragraph{OpenAI o1} Evaluators found the OpenAI o1 model to perform well but to be inconsistent in its citation practices. At times, it provided incorrect citations, while in other cases, its references were remarkably precise. The sometimes exemplary explanations and references let the evaluators independently from each other speculate whether some questions with answers may have appeared in its training data. The model occasionally demonstrated strong reasoning and was typically able to cite the correct EPC as well as PCT rules and articles with a high degree of accuracy, even down to specific sentence fragments or list items. It was able to analyze and interpret even a few complicated questions with tricky wording, which included a pitfall, when their text was very close to phrases in EPC articles. Its explanations tended to be much more detailed than those in the brief answer keys published by the EPO.\footnote{There are more extensive solutions and commentaries available commercially for candidates who prepare for the exam. However, they may likely not be part of the pre-training data as they are typically not readily available online.}
 
Despite the positive overall impression, o1 was criticized in detail on rather fundamental points. The model still struggled to separate the rather fundamental concept of novelty from obviousness or inventive step in some situations and also cited wrong articles or wrong paragraphs within the right article in such cases. Among the issues was, for example, the case of unpublished European patent documents within their 18-month period, which the model also wanted to exclude from novelty. The model did not apply nor explain correctly that novelty does not need features that had never been published before but only that no document published all features together. The failures differed from GPT-4o, which rather struggled with the combination of critical prior-art documents with priorities.
 
Overall, the answers and evaluation comments from the patent professionals indicate an attempt at thoroughness, though not always accuracy.

\paragraph{DeepSeek v3}
DeepSeek v3 practically never cited any regulations. It struggled from time to time with sentences that contained many conditions in if--then logics and could either not subsume correctly or missed the logical connection of the requirements (e.g., mandatory to fulfill all vs.\ only one). In various cases, the questions---which were often phrased as statements---appeared to be taken by the model as fact so that they replaced regulation facts. Mistral appeared to demonstrate a similar behavior.

More importantly, DeepSeek v3 missed more intricate novelty--inventiveness rules in combination with priorities and\,/\,or not yet published prior patent applications. The model, for instance, missed the point that patent applications have to be novel over earlier filed but not yet published earlier applications but not inventive. DeepSeek v3 however could apply the novelty concept well and differentiate it from novelty if confronted very specifically with combinations of feature sets that were novel but not necessarily inventive. It demonstrated similar weaknesses in regulations about divisional applications, which it shared with other models (e.g., GPT4o, Gemma 2, or Llama 8B), although some models (e.g., Opus, Mistral Large, Llama 70B, OpenAI o1, and also DeepSeek R1) managed them. However, it was not clear if wrong answers and explanations were caused by \textit{knowledge gaps}, i.e., missing information in the training data, or an issue with language logic. The latter would align with the observation of failure to deal with more intricate logic statements. 
However, DeepSeek v3 certainly had exposure to PCT and\,/\,or EPC regulations or secondary literature about them in its training repertoire as it could answer some pure knowledge questions that were too specific to hallucinate them correctly or take them from other jurisdictions.

The model also struggled with deadline calculations when at least one time was only given roughly (e.g., month or season names) instead of an exact date. Except for Llama 8B, most other models appeared to deal better with such cases. DeepSeek v3 also miscalculated deadlines where the question provided exact dates.

Overall, the explanations were typically very brief and compact, which impeded a further analysis. 
In a few cases, such single-sentence answers were very close to the short answers provided by the EPO, while widely wrong answers in many other questions would not support the hypothesis that the entire set of brief answers was in the training set.

\paragraph{DeepSeek R1}
The DeepSeek reasoning model performed very differently according to the experts' observations. The model still does not often cite references but when it does, they were mostly right. The answers are not as recitent as v3 but sometimes delivered page-long, highly colloquial (``This is tricky'', ``Wait, no'') texts that between dialogues and streams of consciousness. With the information not given to the evaluators that these answers belonged to a reasoning model, clarifies that many answers appear to present the often hard to read raw output stream in which statements are iteratively processed and combined. The model from time to time speculated and hypothetically expanded cases but in contrast to hallucinations of other models clearly stated those aspects subjunctively.

It could answer and correctly explain several cases, where DeepSeek v3 failed. Specifically, as the patent experts commented on it in both cases, the regulations around European divisional applications, their priorities, and the limitation to the original disclosure.

The calculation skills of deadlines and fees were mostly right, but missed holidays. DeepSeek R1 often provided justification or breakdowns for calculations that allowed a deeper look and in various cases accounted for special cases, such as delivery aspects (e.g., Rule 126 EPC), though not always unambiguously correct. It missed holidays though and has shown cases where it speculated in subjunctive form that a date might fall onto a weekend, which could move the deadline to the next business day instead of knowing it for the specific date.

\paragraph{Gemma 2} Finally, the Gemma 2 model often produced incorrect explanations and either did not include references (the majority of cases) or entirely hallucinated wrong ones, though from the right documents. In a few cases, where the referenced article was on the right topic, the content Gemma associated with it was wrong.  Gemma's reasoning was generally considered poor. The human expert evaluators suggested that its training data must have clearly included some patent-relevant texts such as the PCT Regulations and the EPC or documents about them. Still, it could at best cite simpler definitions (such as of novelty) but failed in their application.

It frequently miscalculated deadlines but was occasionally also able to identify certain non-obvious dates correctly. Though it struggled to differentiate between novelty and obviousness, evaluators noted that the model appeared to have the correct definitions encoded in its training, as for instance seen in Question 2012--Q5.

\paragraph{Synopsys}
In summary, these evaluation remarks of the patent experts underscore that even models with comparable levels of raw accuracy demonstrate significant disparities in core legal reasoning skills.
 These include differences in the ability to compute deadlines, accurately cite statutory authorities, apply legal rules to a case and given facts, distinguish between novelty and obviousness,
 and deliver clear and persuasive justifications. 
 
 Particularly, the inability to fully grasp and apply the concepts of novelty and obviousness is grave from the perspective of the patent experts. Some models mixed up the terms or even incorrectly equated them (e.g., if (not) inventive then also (not) novel), despite, in many cases, their ability to properly define them when asked directly. Related questions, such as 2012--Q5 or 2015--Q8, practically served here as a litmus test.
 
 Most models struggled with the practical application of aspects of novelty and their different implications, e.g., if other applications with some or all features were filed less than 18 months before so that they are unpublished yet. However, Articles 54\,(3) and 56 EPC (with its explicit reference to 54\,(3)) are very clear and leave no room for interpretation so that some models, such as Opus, managed that with spot-on explanation in an apparently almost effort-less way with correct references to the patent code. Others, among them also GPT-4o and o1, could not and in the few cases where they provided more or less coincidentally the right \textit{True}\,/\,\textit{False} answer, they often delivered rather wrong explanations. Also, Llama 3.1 405B struggled with the combination of novelty, inventiveness, the 18-month period, and priorities. In combination with applications, where only some claims could use an earlier priority (e.g., 2015--Q8), Opus also failed to grasp that some claims in the same patent application are subject to a different prior art.
 
 The shortcomings with novelty, inventive step, and obviousness may also reflect the conflict of legal and everyday language and the high level of language precision particularly needed in patent applications. After all, most LLMs have been trained on corpora with a large share of even the lowest-quality text from the internet. These internet sources may include online discussions and marketing language, which typically do not aim at language precision. Furthermore, the embeddings determine whether 
 similarity may lead to a low model-intrinsic distance between novel and inventive so that LLMs may treat them almost as synonyms. Since those embeddings are rather deeply implanted into language transformers, fine-tuning or other tricks may not be able to solve such issues easily.

\paragraph{Aspect of Embeddings}
 To further investigate the underlying cause for the difficulties observed among language models regarding novelty, obviousness, and inventiveness as well as previous observations that fine-tuning of small models for patent tasks can face limitations, we analyzed embedding similarities across three distinct models: Llama 3.1 (8B), Llama 3.1 (70B), and Gemma 2 (9B). We included both relevant terms as well as irrelevant control terms.
 
 The heatmaps generated from this embedding analysis revealed noteworthy insights. All three models exhibited  embedding
 similarities between terms such as \textit{novel}, \textit{inventive}, \textit{nonobvious}, and \textit{obvious}, which suggest that these terms are associated in their internal representations. 

\begin{figure}[H]
\centering
\includegraphics[width=\linewidth]{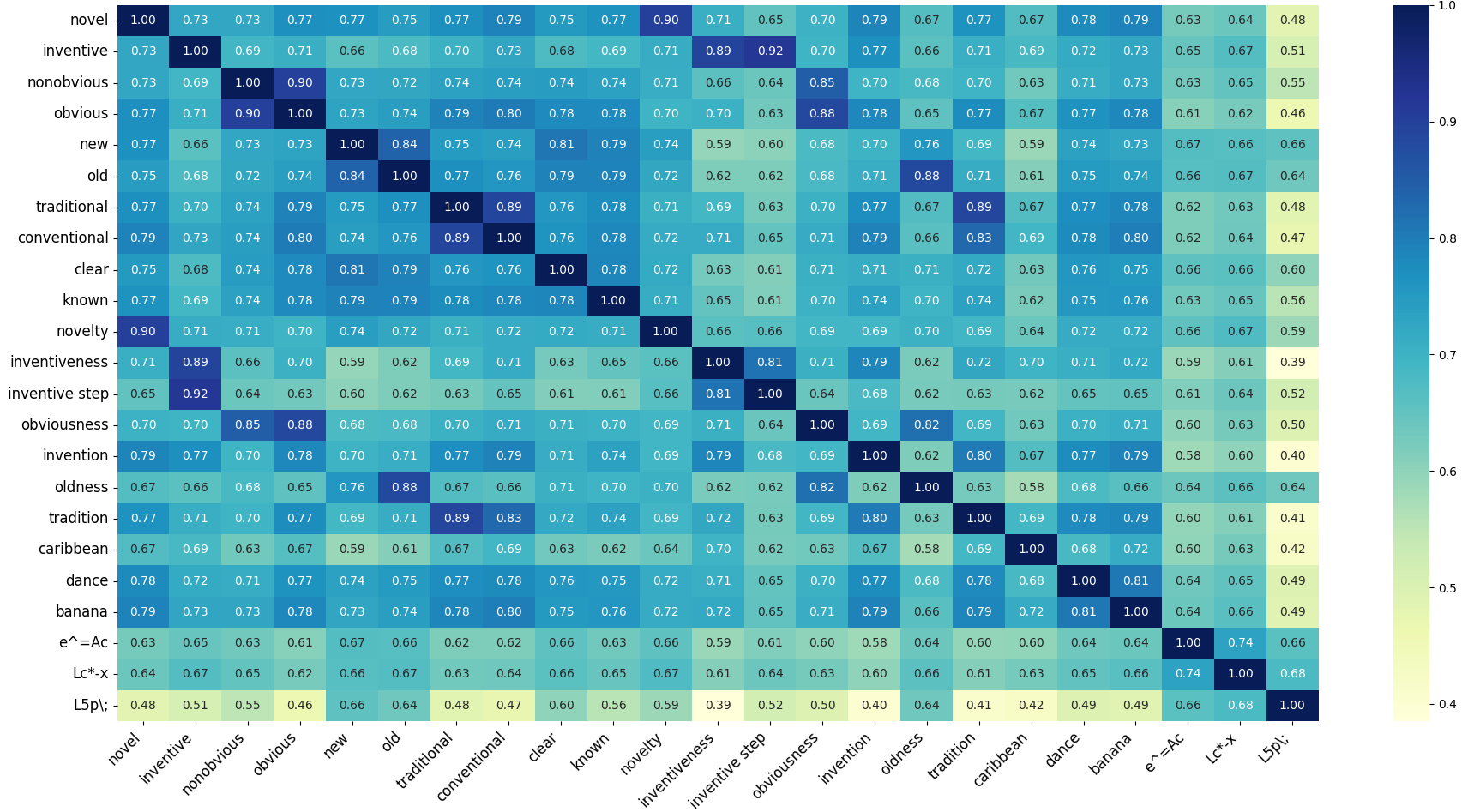}
\caption{Embedding similarity of Gemma 2 9B.}
\label{fig:gemma2}
\end{figure}

 In Gemma 2 (Fig.\ \ref{fig:gemma2}), there is a high embedding similarity between \textit{novel}, \textit{inventive}, \textit{nonobvious}, and \textit{obvious}.
 For example, the similarity between \textit{novel} and \textit{inventive} is $0.73$, which is not so much smaller than between the correctly related \textit{inventive} and \textit{inventive step} ($0.92$) as well as larger than the likewise synonymous \textit{inventive} and \textit{nonobvious} ($0.69$). Similarly, high correlations are seen across many pairs of core legal terms. This high similarity suggests a potential source for confusion between otherwise distinct legal
 concepts, as the model may not create sufficient semantic distance between them.

To validate the specificity of these embeddings, we also incorporated non-related terms such as ``banana", ``caribbean", and rather arbitrary ASCII sequences (e.g., ``\texttt{e\textasciicircum=Ac}", ``\texttt{Lc$^*$-x}", and ``\texttt{L5p\textbackslash;}").\footnote{Note that the models tokenize those strings, and even random letters can share tokens with other terms, though typically not multiple dimensions of the embedding vector, which represents several tokens. The rather arbitrary ASCII symbols should reduce such spurious correlations.}  
These unrelated control terms exhibited lower embedding similarity to the legal terms.

Thus, the high similarity observed between legal terms in some models may indeed reflect insufficient semantic distance due to the inappropriate training data and indicate a source for confusion. The embedding and the training data could be a reason why even advanced language models struggle with subtle yet crucial legal distinctions. Part of the struggle may be linked to embeddings formed during pretraining on general internet data, which follow everyday interpretations of \textit{novel} and \textit{inventive}. The strong semantic proximity between legally distinct terms in some of the models emphasizes the difficulty in refining
 their understanding purely through fine-tuning or prompt engineering alone, which leaves the embedding unchanged. Instead, there is a deeper linguistic entanglement that could require deeper or more sophisticated approaches to resolve.

\subsection{Comparative Analysis of GPT-4o and Opus}
In most years (particularly 2012, 2014, and 2017), GPT-4o matched or outperformed Opus in many of the measures captured—including accuracy, precision, F1 score, BLEU, ROUGE-1, ROUGE-L, and BERTScores, though not in recall.

GPT-4o answered more accurately in both 2012 ($0.8$ for GPT-4o vs.\! $0.67$ for Opus) and 2014 ($0.65$ vs.\! $0.61$). While both models scored $0.71$ in 2017, GPT-4o fared better in every other metric.
The accuracy comparison may suggest a general performance advantage of GPT-4o over Opus.

GPT-4o also dominated in precision. It had a consistent edge over Opus—specifically in precision (for instance, $0.58$ vs.\! $0.57$ in 2014 and $0.67$ vs.\! $0.59$ in 2017).

However, Opus performed better than GPT-4o with respect to recall ($0.79$ vs.\! $0.89$ in 2012, $0.61$ vs.\! $0.78$ in 2014, and $0.69$ vs.\! $1$ in 2017).

The language scores reflect a closer head-to-head race. In 2012, GPT-4o scored a BLEU of $0.16$ in 2012 (compared to Opus’ $0.11$), ROUGE-1 of $0.73$ (compared to $0.60$), ROUGE-L of $0.30$ (compared to $0.27$), and BERTScore of $0.78$ (compared to $0.77$). In 2014, GPT-4o had a slightly lower BLEU score of $0.08$ (compared to $0.09$), ROUGE-1 of $0.65$ (also $0.65$), ROUGE-L of $0.28$ (compared to $0.27$), and BERTScore of $0.84$ (compared to $0.83$). Also in 2017, GPT-4o had a higher BLEU ($0.13$ compared to Opus’ $0.05$), ROUGE-1 ($0.69$ compared to $0.54$), and BERTScores ($0.84$ compared to $0.82$).
However, readers need to mind that the metrics beyond the logical comparison with the ground truth (accuracy) are mostly language qualities or self-referenced features that typically do not matter for correctness or legal precision.

On the evaluators' side, GPT-4o received slightly favorable reviews compared to Opus or was at least close. Although GPT-4o fell slightly behind Opus in 2012 ($0.57$ vs.\! $0.58$), it was ahead in both 2014 ($0.55$ vs.\! $0.48$) and 2017 ($0.49$ vs.\! $0.45$).
Although both models cited primary sources only rarely and often inaccurately, reviewers found GPT-4o’s explanations were a little more coherent, more readable, and contained somewhat more citations than
 Opus’s, even in situations in which raw quality measures between the two models were essentially the same.

However, GPT-4o's struggles with the application of fundamental concepts such as novelty and inventive steps, as well as patentability, are a rather substantial concern, which may outweigh any overall higher accuracy score. The results may underline that automated metrics can provide some insights, but they do not pick up the content, clarity, correct logic, logical structure, or quality of citations.  

\subsection{Human Vs. Machine}

The professional human evaluation provided numerical scores (from 1 to 10 linearly scaled to 0 to 1 in postprocessing) on individual questions for conceptual clarity, coherence, and justification quality beyond mere correctness. Upon closer analysis, it was observed that models such as Sonnet 3.5 and Python Llama 3.1 8B  negatively correlated with correctness (respectively $-0.07$ and $-0.16$). Thus, evaluators awarded scores more for justification quality and logical consistency rather than solely on the \textit{True} or \textit{False} answers so that random guessing could not score points easily, and correct reasoning, fact extraction from the questions, and subsumption were necessary. This observation underscores the nuanced nature of human evaluation and highlights the critical role that qualitative factors play in expert decision-making processes. Thus, human oversight remains essential, as computational accuracy alone does not fully encapsulate expert-level understanding or reasoning.

\section{Conclusion}

Thus, this study underscores both the capabilities and limitations of large language models (LLMs) in specialized legal domains, particularly patent law. Whereas larger models demonstrate considerable promise with high accuracy in multiple-choice and structured legal contexts, significant limitations remain in tasks that require deeper interpretive reasoning, comprehensive legal analysis, and logical consistency. Although OpenAI o1 and Sonnet 3.5 have shown substantial alignment with human evaluators, particularly in their ability to provide logically consistent and coherent justifications—the mean accuracy still never exceeded 90\%.
No model below 100B degrees of freedom exceeded the odds of pure guessing (accuracy of 0.5). Some models even fell far behind this bar (e.g., Llama3.1 8B with 0.39 and Gemma 2 with 0.37) so that a simple coin toss would have been superior.

Smaller models, furthermore, consistently struggled with the nuanced and indirect implications of patent law. Some models may struggle already on the level of their word embedding, which could be too closely linked to the everyday interpretation of terms. As the embedding is formed during pre-training, usually, a mismatch or bias could indicate that fine-tuning of such models to patent language has a massive bottleneck and may explain reported struggles in the literature.

Overall, the results suggest that the depth and consistency required for comprehensive patent examination remain out of reach for current AI models.

\newpage
\begin{appendices}
\counterwithout{figure}{section} 
\renewcommand{\thefigure}{\arabic{figure}} 
\setcounter{figure}{7} 
\section{Model‑by‑model answer accuracy}\label{secA}

\begin{figure}[H]
\centering
\includegraphics[width=0.95\linewidth]{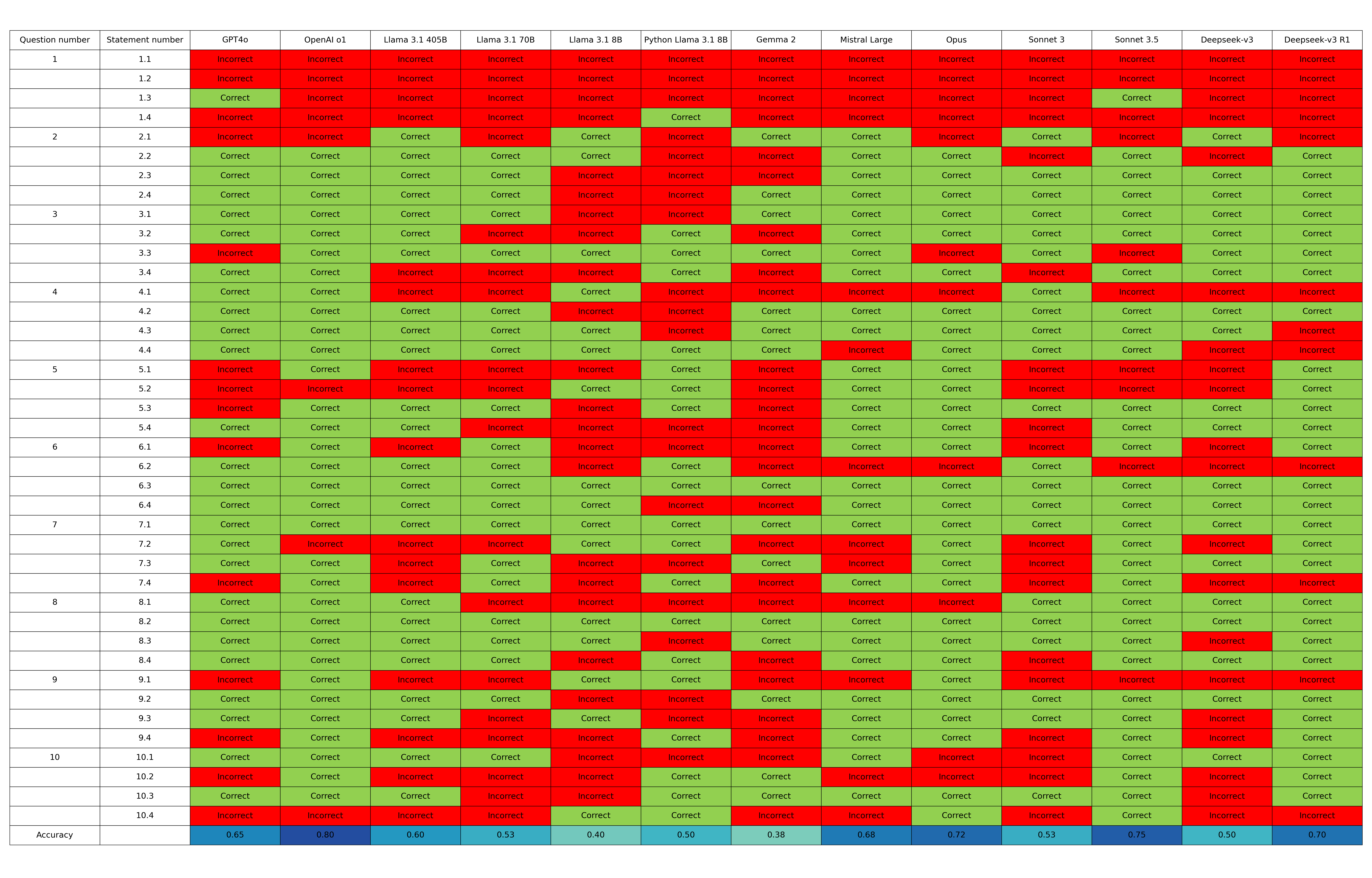}
\caption{Model‑by‑model answer accuracy for the year 2012.}
\label{fig:2012}
\end{figure}

\begin{figure}[H]
\centering
\includegraphics[width=0.95\linewidth]{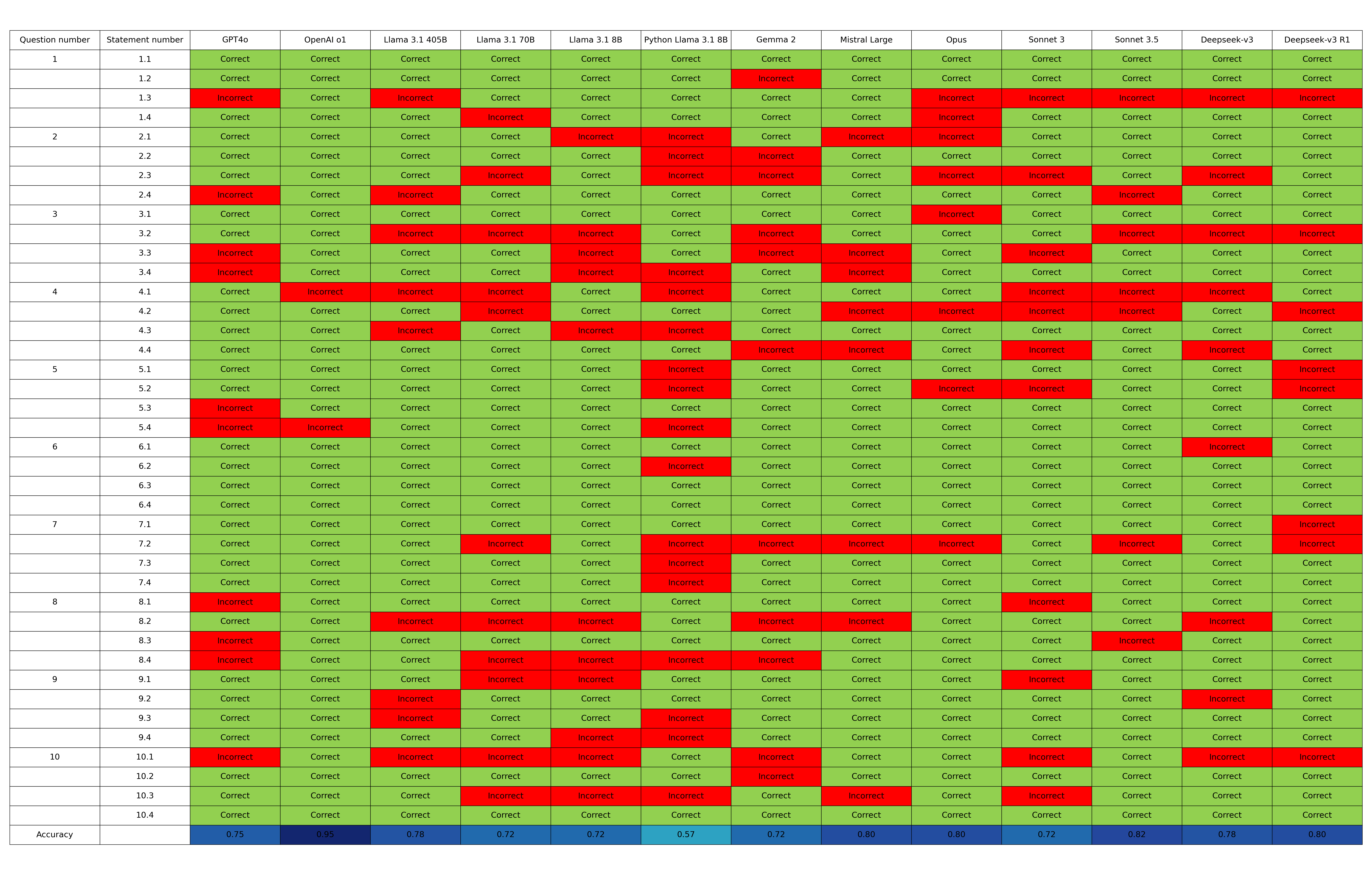}
\caption{Model‑by‑model answer accuracy for the year 2013.}
\label{fig:2013}
\end{figure}

\begin{figure}[H]
\centering
\includegraphics[width=0.95\linewidth]{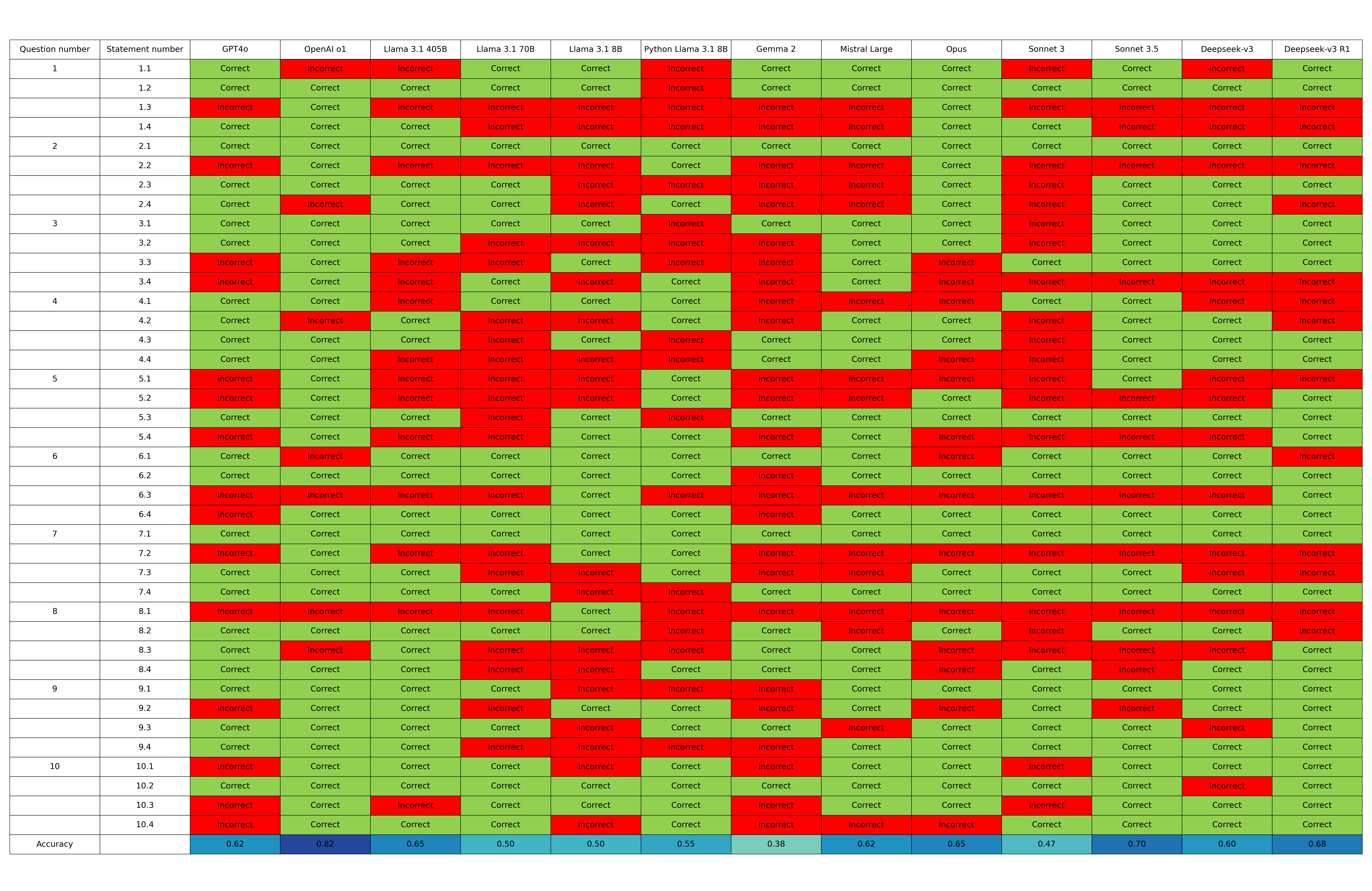}
\caption{Model‑by‑model answer accuracy for the year 2014.}
\label{fig:2014}
\end{figure}

\begin{figure}[H]
\centering
\includegraphics[width=0.95\linewidth]{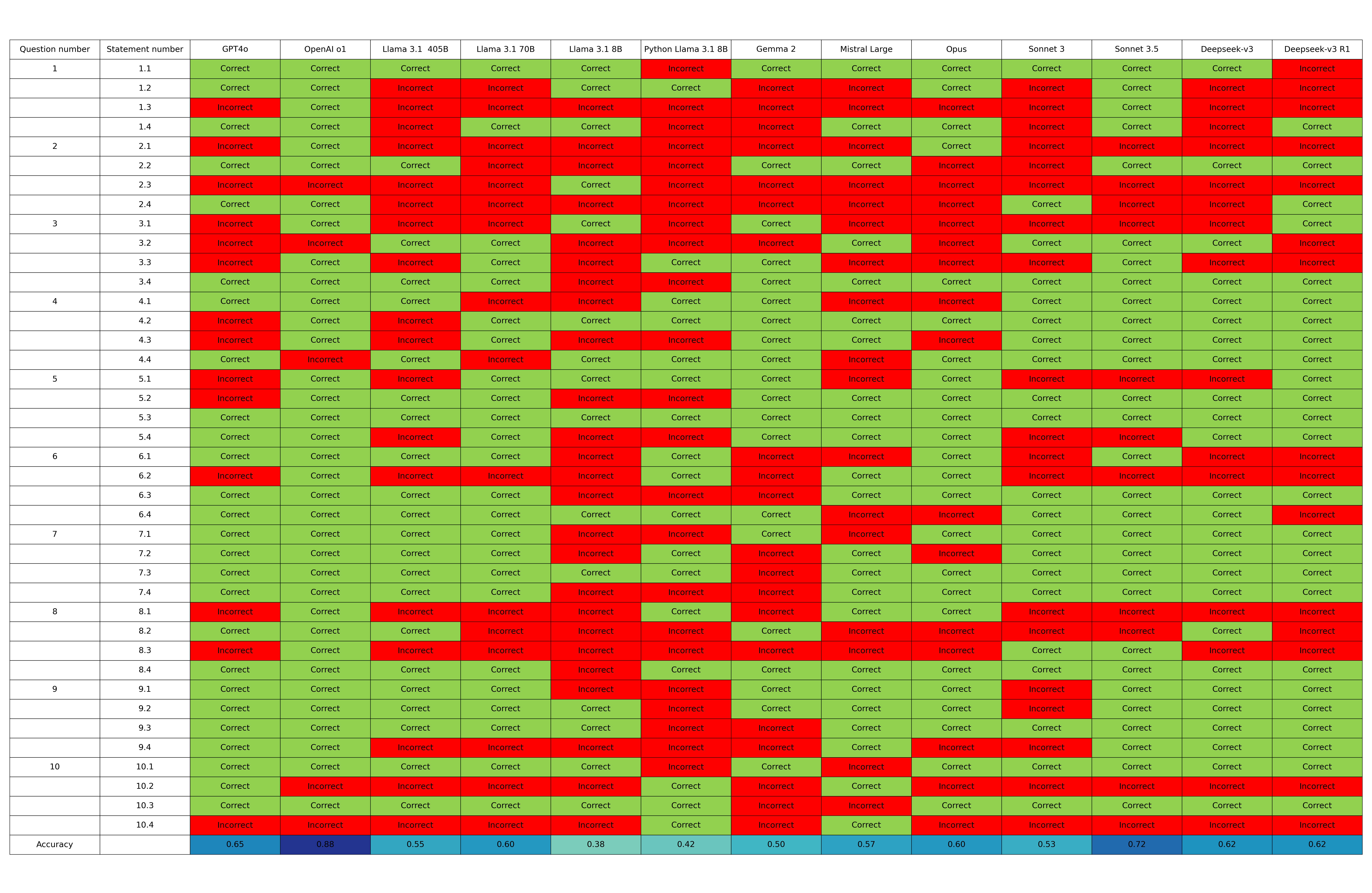}
\caption{Model‑by‑model answer accuracy for the year 2015.}
\label{fig:2015}
\end{figure}

\begin{figure}[H]
\centering
\includegraphics[width=0.95\linewidth]{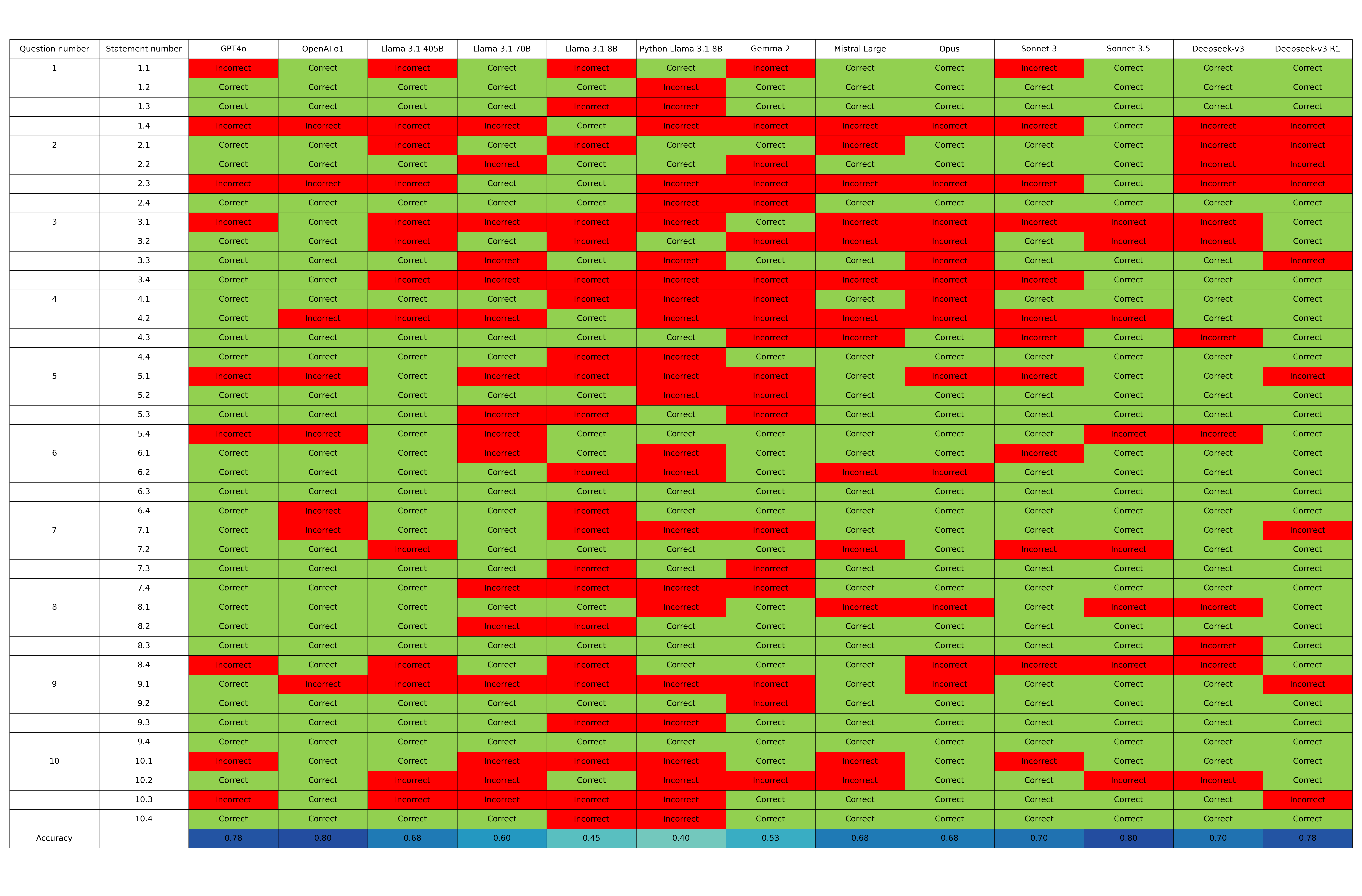}
\caption{Model‑by‑model answer accuracy for the year 2016.}
\label{fig:2016}
\end{figure}

\begin{figure}[H]
\centering
\includegraphics[width=0.95\linewidth]{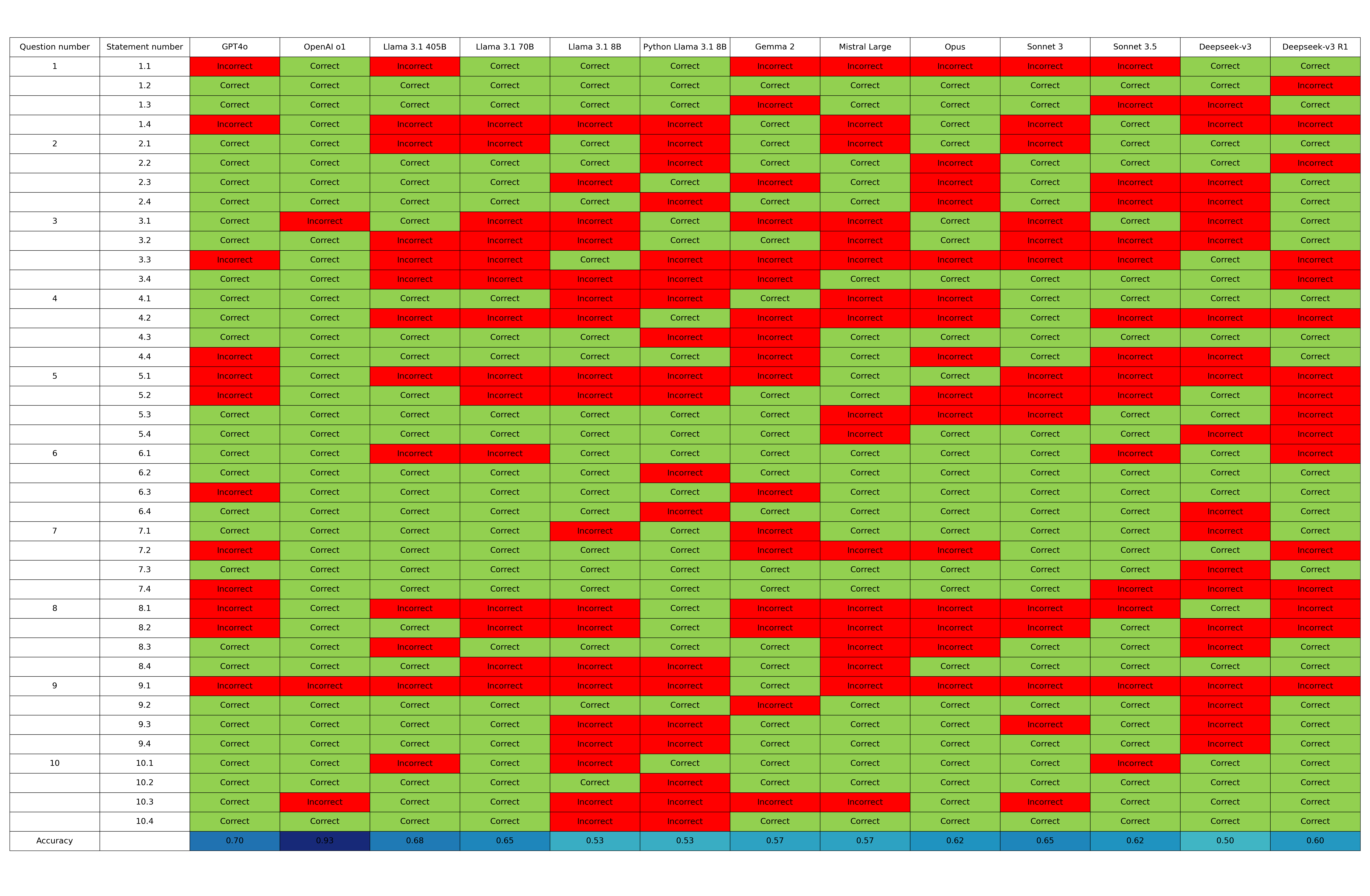}
\caption{Model‑by‑model answer accuracy for the year 2017.}
\label{fig:2017}
\end{figure}

\begin{figure}[H]
\centering
\includegraphics[width=0.95\linewidth]{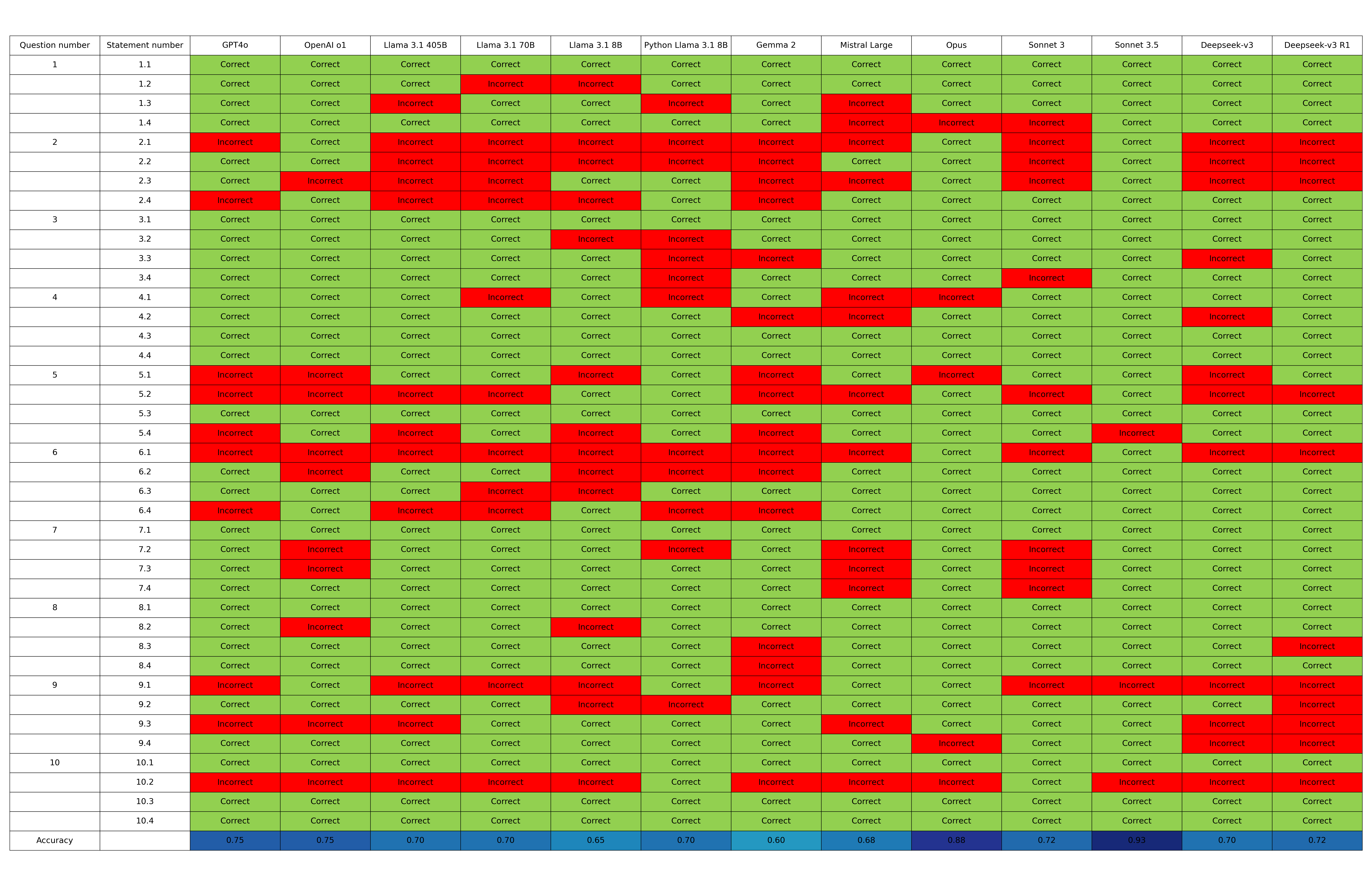}
\caption{Model‑by‑model answer accuracy for the year 2018.}
\label{fig:2018}
\end{figure}

\begin{figure}[H]
\centering
\includegraphics[width=0.95\linewidth]{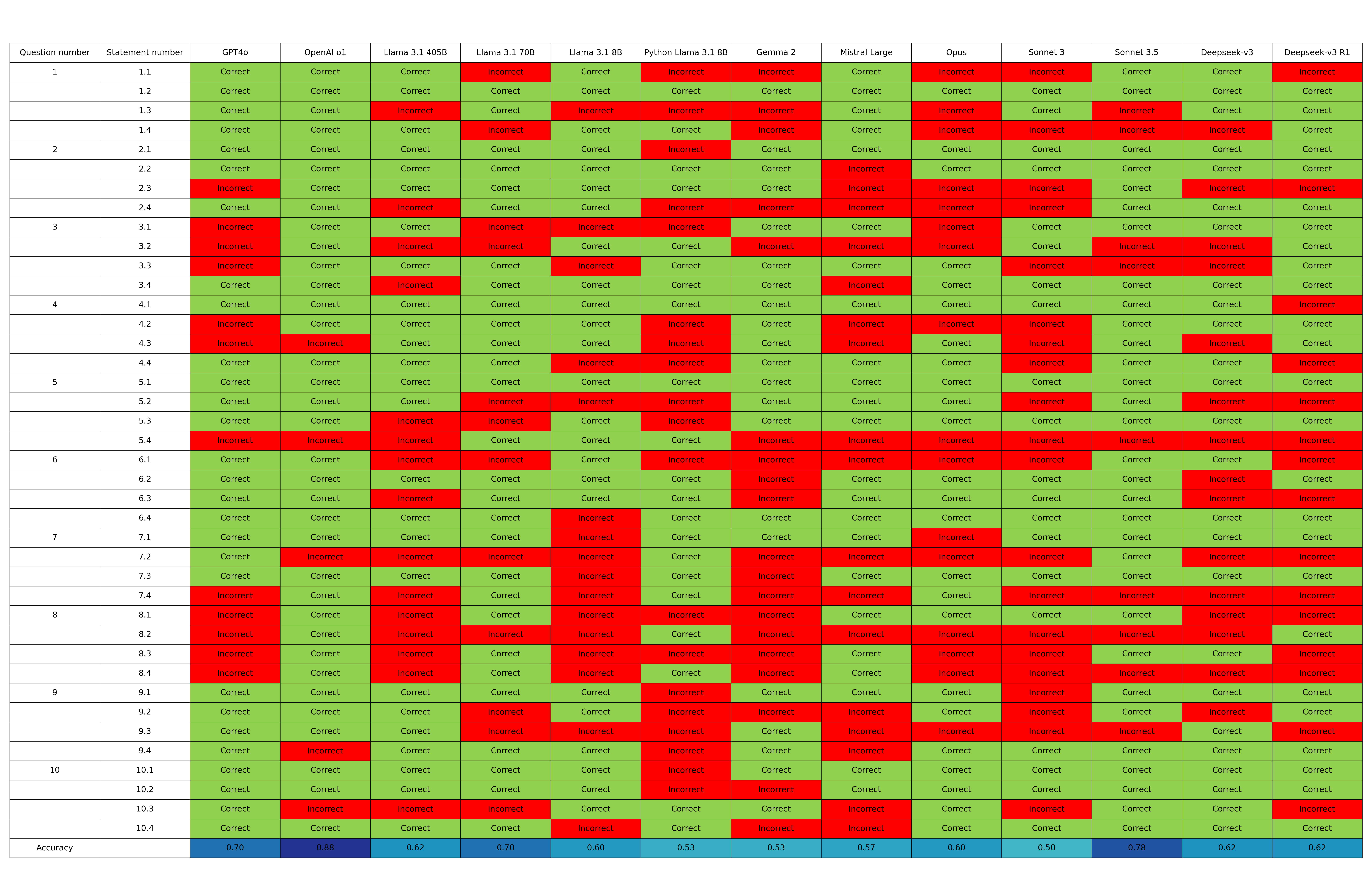}
\caption{Model‑by‑model answer accuracy for the year 2019.}
\label{fig:2019}
\end{figure}

\begin{figure}[H]
\centering
\includegraphics[width=0.95\linewidth]{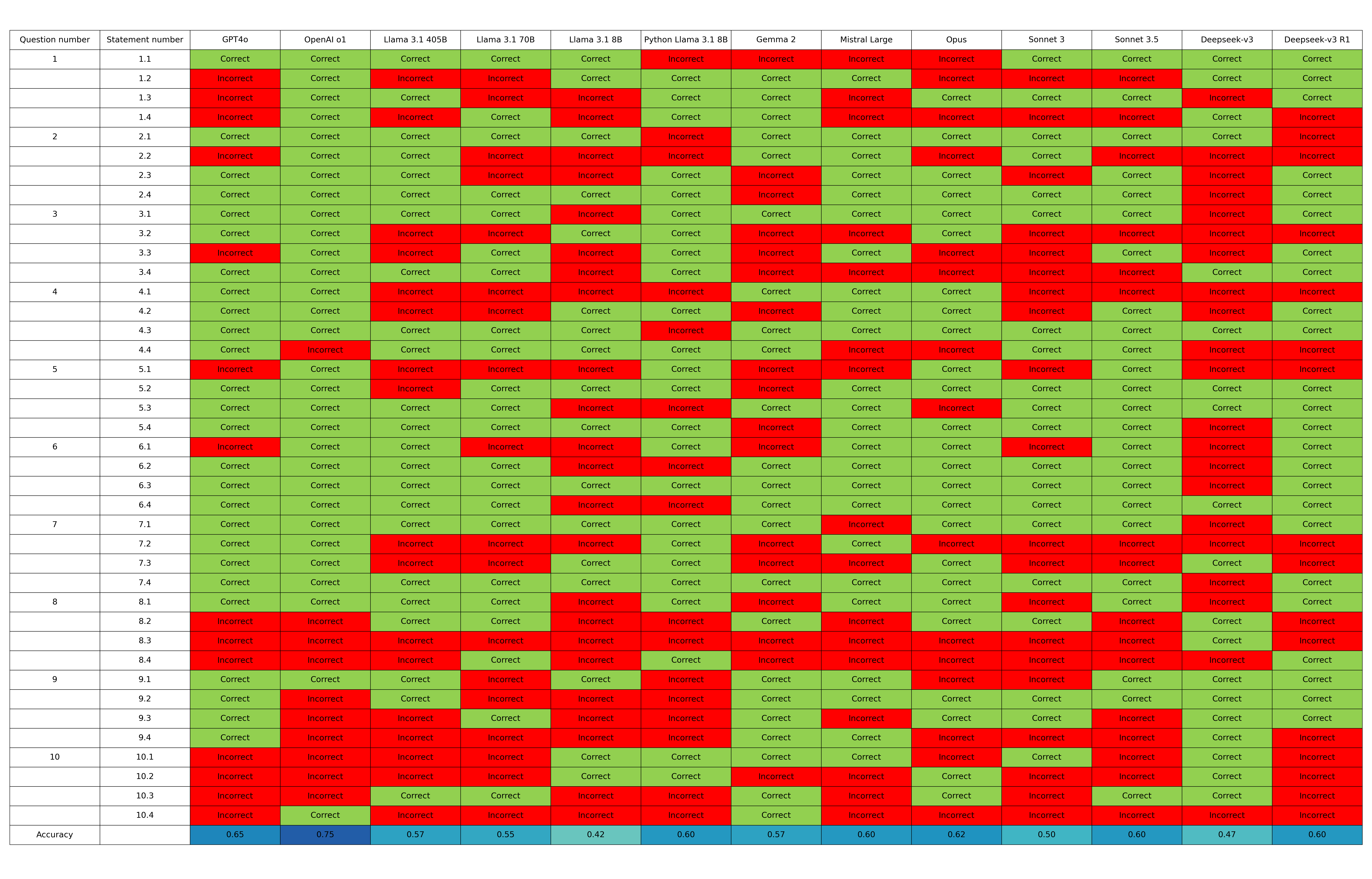}
\caption{Model‑by‑model answer accuracy for the year 2021}
\label{fig:2021}
\end{figure}

\begin{figure}[H]
\centering
\includegraphics[width=0.95\linewidth]{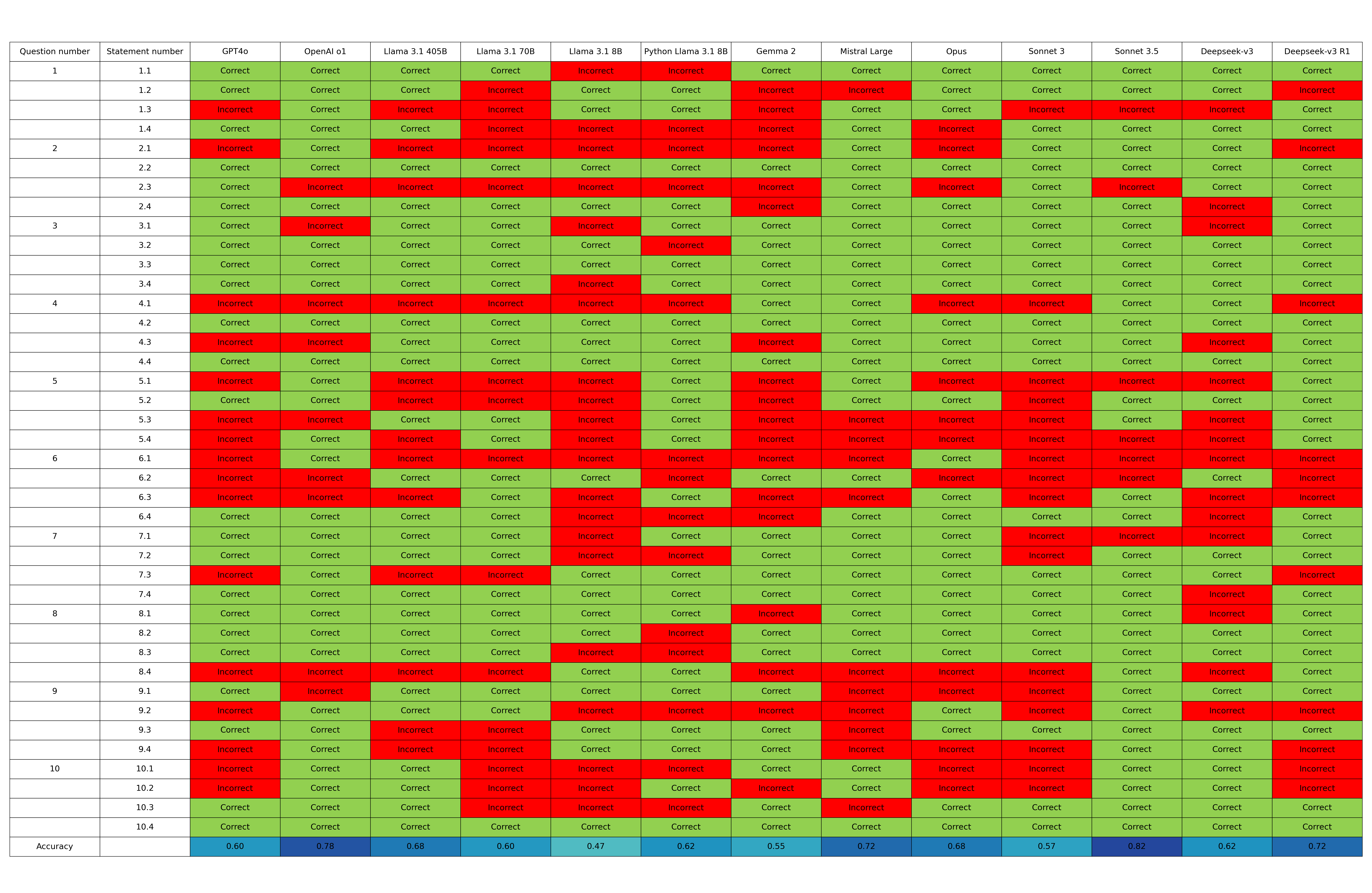}
\caption{Model‑by‑model answer accuracy for the year 2018.}
\label{fig:2022}
\end{figure}

\begin{figure}[H]
\centering
\includegraphics[width=0.95\linewidth]{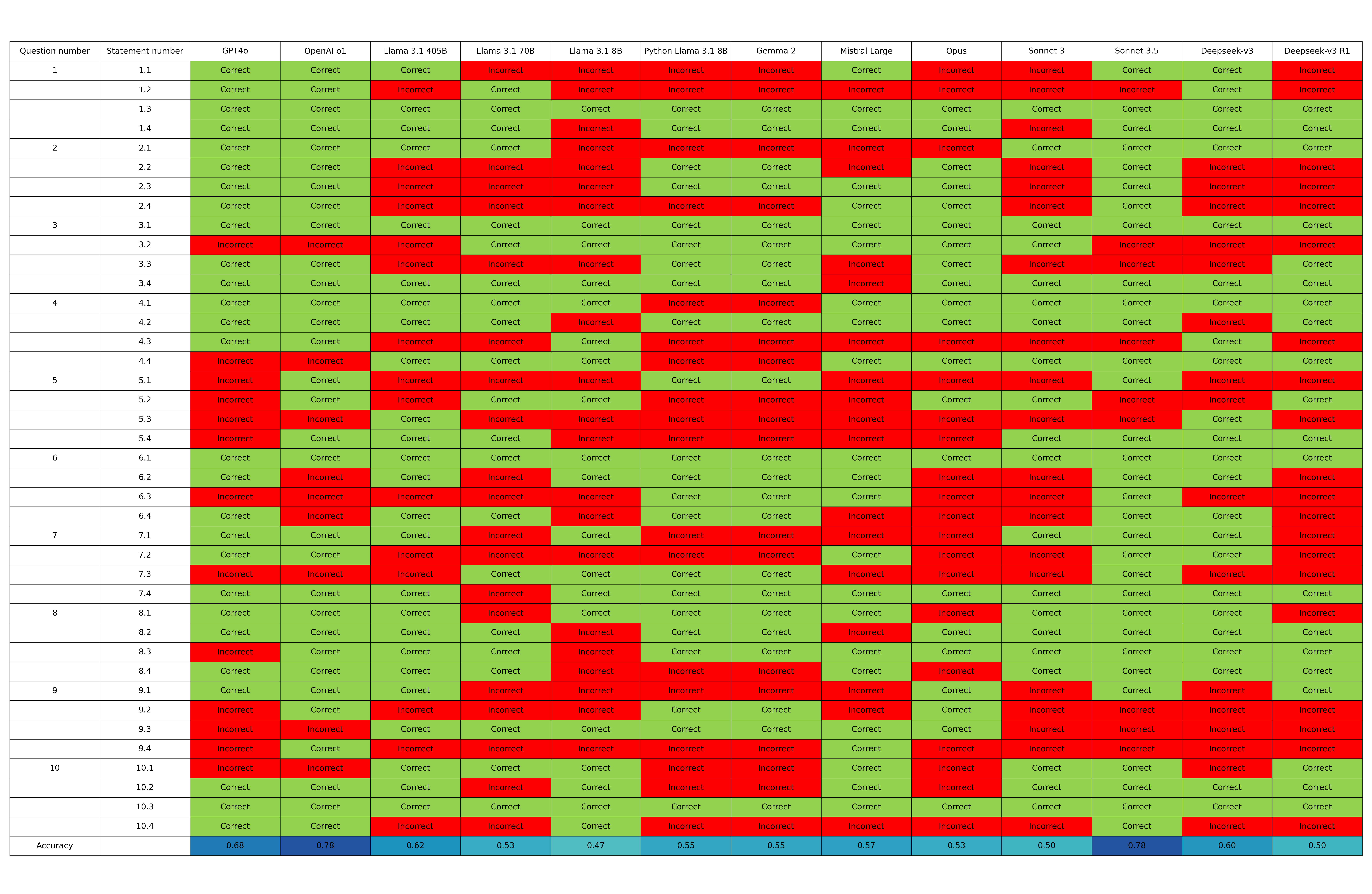}
\caption{Model‑by‑model answer accuracy for the year 2023.}
\label{fig:2023}
\end{figure}

\begin{figure}[H]
\centering
\includegraphics[width=0.95\linewidth]{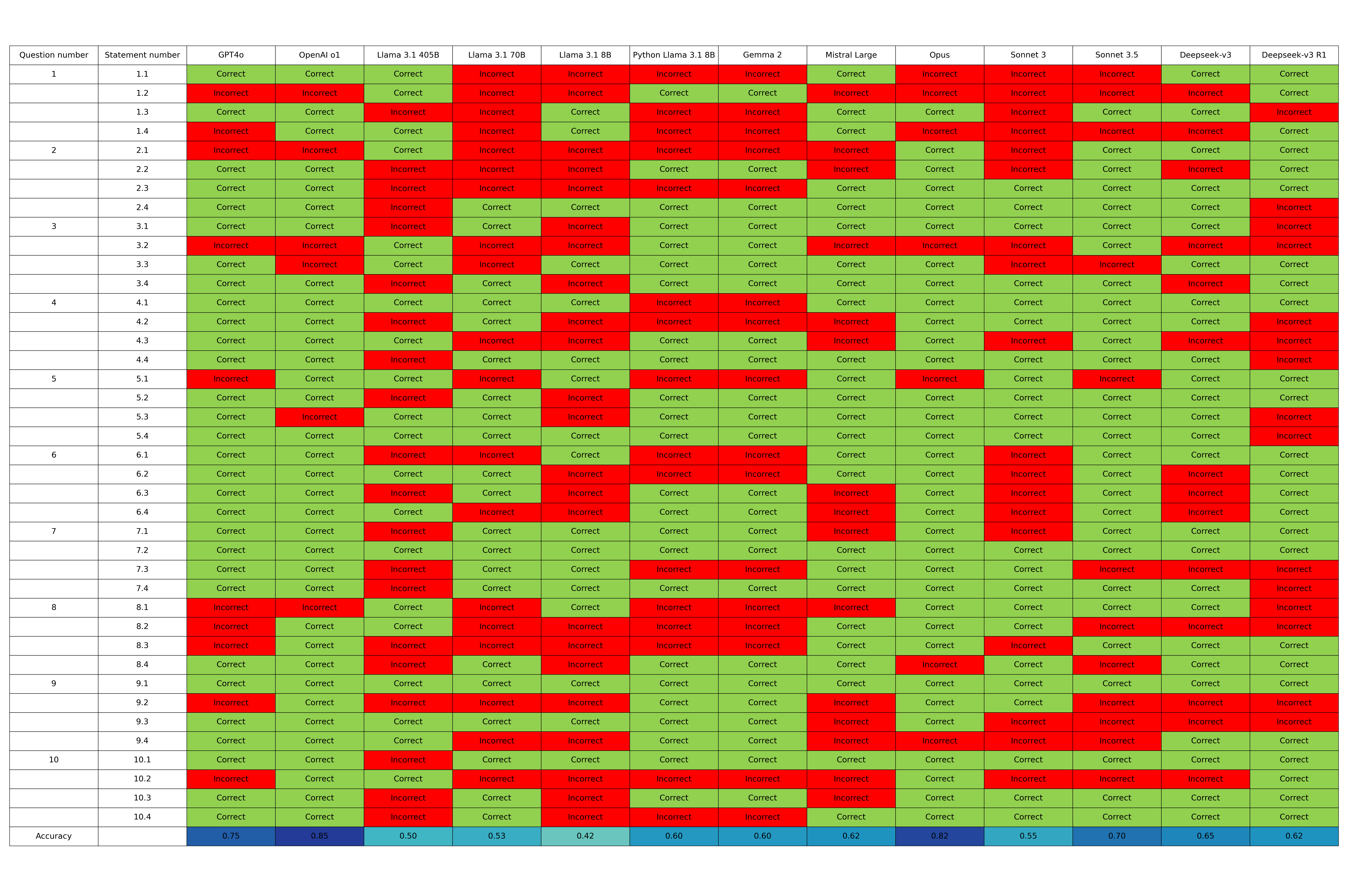}
\caption{Model‑by‑model answer accuracy for the year 2024.}
\label{fig:2024}
\end{figure}

\newpage

\section{Embedding similarity}\label{secB}
\begin{figure}[H]
\centering
\includegraphics[width=\linewidth]{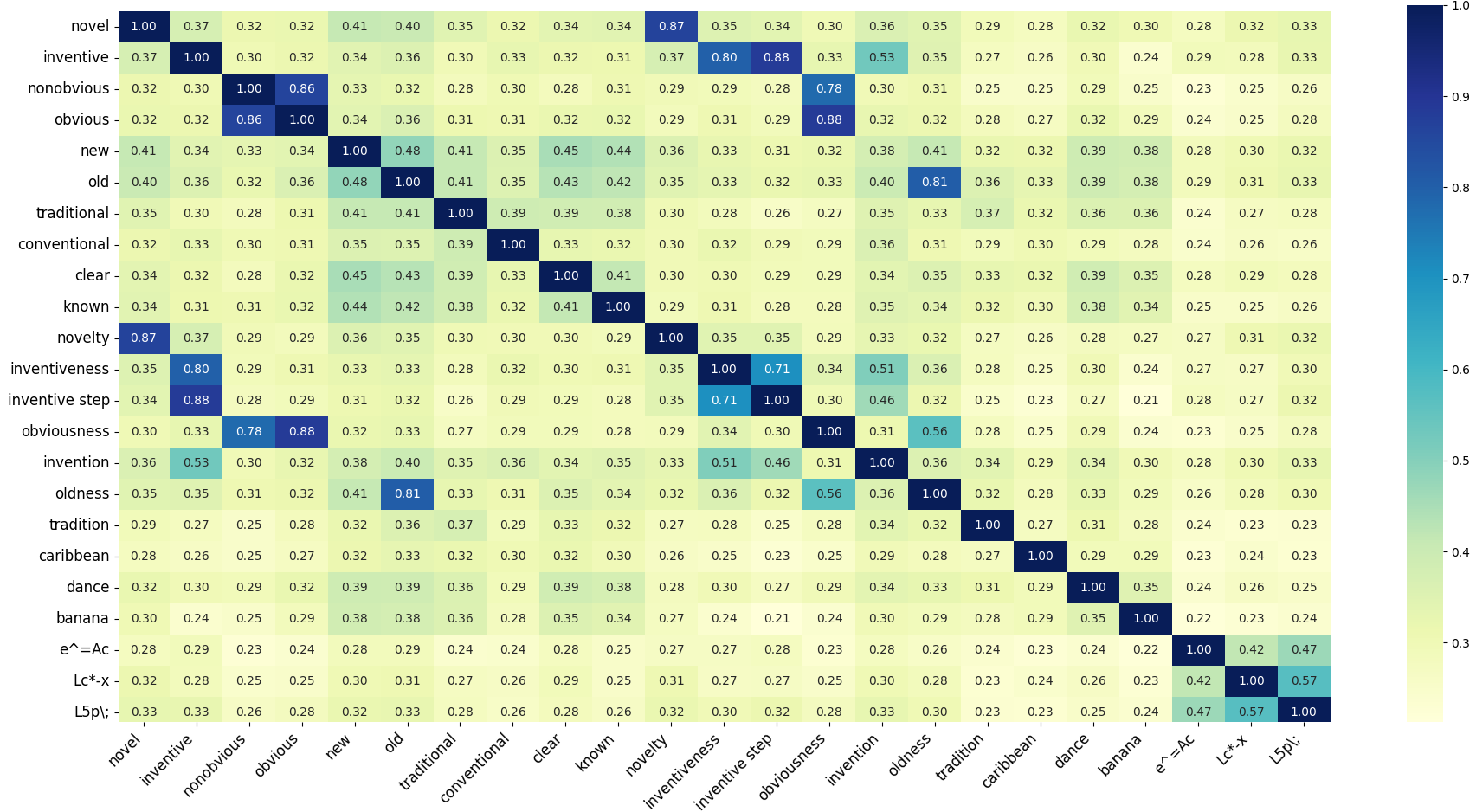}
\caption{Embedding similarity of Llama 3.1 8B.}
\label{fig:Llama8b}
\end{figure}

\begin{figure}[H]
\centering
\includegraphics[width=\linewidth]{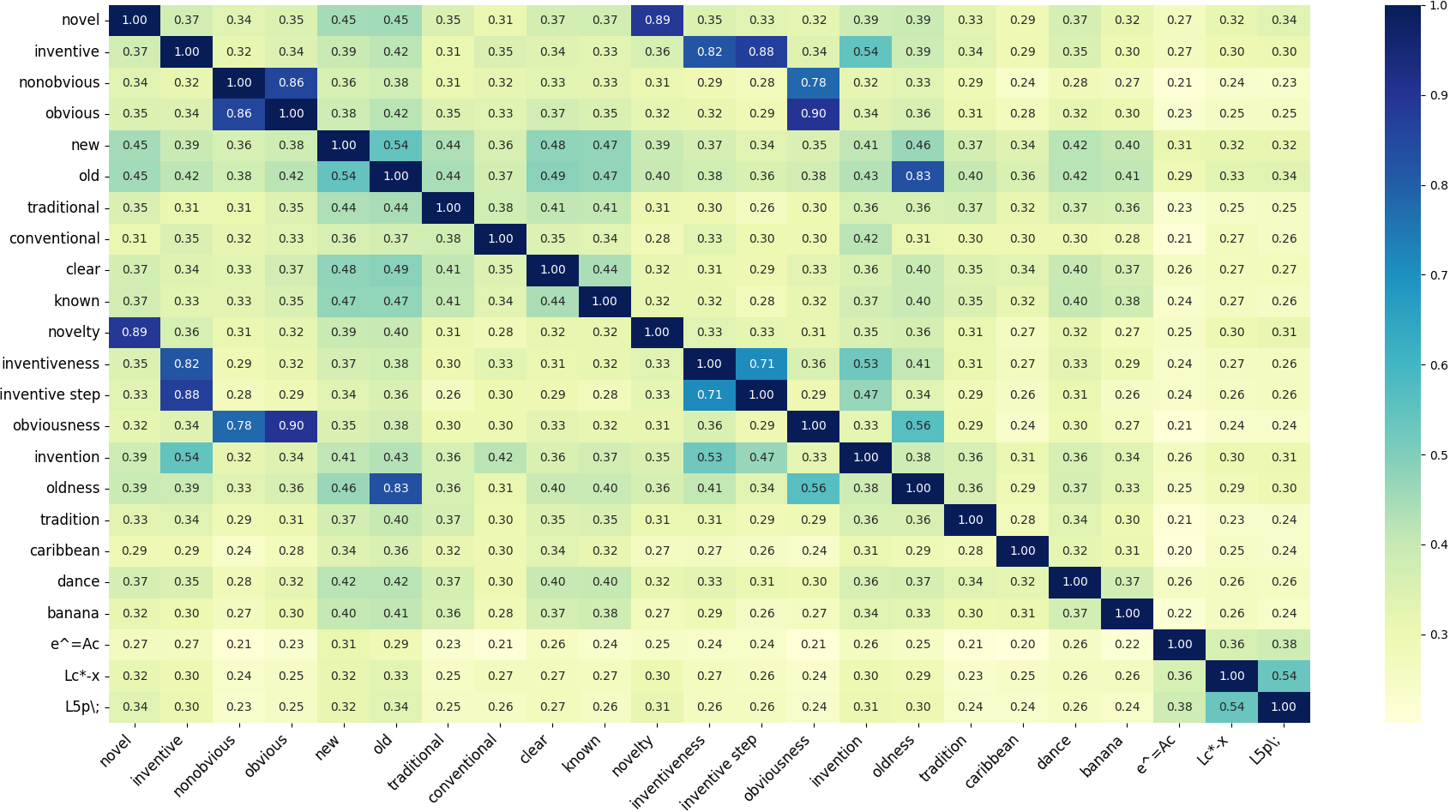}
\caption{Embedding similarity of Llama 3.1 70B.}
\label{fig:Llama70b}
\end{figure}

\end{appendices}
\newpage

\bibliography{references_out}

\end{document}